\documentclass[conference]{IEEEtran}
\usepackage[dvipsnames, svgnames, x11names]{xcolor}

\IEEEoverridecommandlockouts
\usepackage{cite}
\usepackage{amsmath,amssymb,amsfonts}
\usepackage[ruled,linesnumbered]{algorithm2e}

\SetCommentSty{mycommfont}

\usepackage{graphicx}
\usepackage{textcomp}
\usepackage{xcolor}
\usepackage{comment}
\usepackage[numbers]{natbib}

\usepackage{braket}
\usepackage{hyperref}
\usepackage{float}
\usepackage{multirow}
\usepackage{makecell}
\usepackage{bm}
\usepackage{subfig}
\usepackage{setspace}
\usepackage{booktabs}

\usepackage{algorithmic}
\usepackage[breakable]{tcolorbox}
\usepackage{tagging}
\usepackage[ampersand]{easylist}
\usepackage{amssymb}
\usepackage{listings}
\usepackage{xurl}

\usepackage{sc} 

\def\BibTeX{{\rm B\kern-.05em{\sc i\kern-.025em b}\kern-.08em
    T\kern-.1667em\lower.7ex\hbox{E}\kern-.125emX}}
\begin{document}

\makeatletter
\newcommand{\linebreakand}{%
  \end{@IEEEauthorhalign}
  \hfill\mbox{}\par
  \mbox{}\hfill\begin{@IEEEauthorhalign}
}

\title{
Queen: A quick, scalable, and comprehensive quantum circuit simulation for supercomputing\\
{
\footnotesize 
Welcome to compare to our state-of-the-art simulator released in 2024 or offer paid consultation for performance improvement.
}
\thanks{We thank the National Center for High-Performance Computing for providing access to the NVIDIA DGX-A100 workstation.}
}

\makeatother

\author{
\IEEEauthorblockN{Chuan-Chi Wang}
\IEEEauthorblockA{\textit{National Taiwan University} \\
\textit{Taipei, Taiwan}\\
d10922012@ntu.edu.tw
}
\and
\IEEEauthorblockN{Yu-Cheng Lin}
\IEEEauthorblockA{\textit{National Taiwan University} \\
\textit{Taipei, Taiwan}\\
r11922015@csie.ntu.edu.tw
}
\and
\IEEEauthorblockN{Yan-Jie Wang}
\IEEEauthorblockA{\textit{National Taiwan University} \\
\textit{Taipei, Taiwan}\\
yanjiewtw@gmail.com
}
\\ 
\linebreakand 
\IEEEauthorblockN{Chia-Heng Tu}
\IEEEauthorblockA{\textit{National Cheng Kung University} \\
\textit{Tainan, Taiwan}\\
chiaheng@ncku.edu.tw
}
\and 
\IEEEauthorblockN{Shih-Hao Hung}
\IEEEauthorblockA{\textit{National Taiwan University} \\
\textit{Taipei, Taiwan}\\
hungsh@csie.ntu.edu.tw
}
}

\maketitle

\begin{abstract}
The state vector-based simulation offers a convenient approach to developing and validating quantum algorithms with noise-free results. However, limited by the absence of cache-aware implementations and unpolished circuit optimizations, the past simulators were severely constrained in performance, leading to stagnation in quantum computing.
In this paper, we present an innovative quantum circuit simulation toolkit comprising gate optimization and simulation modules to address these performance challenges. For the performance, scalability, and comprehensive evaluation, we conduct a series of particular circuit benchmarks and strong scaling tests on a DGX-A100 workstation and achieve averaging 9 times speedup compared to state-of-the-art simulators, including QuEST, IBM-Aer, and NVIDIA-cuQuantum. Moreover, the critical performance metric FLOPS increases by up to a factor of 8-fold, and arithmetic intensity experiences a remarkable 96x enhancement.
We believe the proposed toolkit paves the way for faster quantum circuit simulations, thereby facilitating the development of novel quantum algorithms.

\end{abstract}



\section{Introduction}
\label{sec:introduction}
Quantum computing is a rapidly developing field, with various types of quantum computers being actively explored and developed. This emerging technology has attracted significant attention because of its revolutionary potential to outperform classical computers in addressing complex problems. It has already begun to influence various domains, such as cryptography, machine learning, and combinatorial optimization problems. 



As the development of quantum computers faces significant challenges, such as noise, scalability, and economic feasibility, quantum circuit simulation (QCS) serves as an important alternative for developing and evaluating novel quantum algorithms on a classical computer. This work primarily focuses on the state vector-based simulation, which is also known as the Schrödinger-style full-state simulation. This simulation scheme stores a full state vector in memory, which is updated following each quantum operation to account for the operation’s impact. It has become one of the most extensively studied and utilized QCS since it is able to simulate an arbitrary form of quantum circuits, easily debug quantum applications, and execute noise-free operations.



The time required by QCS is increased with the number of quantum bits (denoted as qubits) and the size of quantum circuit (the number of quantum gates). Numerous studies have been done to accelerate state vector-based simulations. These studies cover a wide vertical spectrum from high-level quantum circuit optimizations by reducing quantum gate numbers to low-level computer architecture related optimizations by parallelizing QCS. For instance, ongoing research works delve into \emph{gate fusion} techniques~\cite{Cirq, qHiPSTER, 45qubit} and \emph{qubit reordering} strategies~\cite{45qubit, cache_blocking, cpu_gpu_communication} for QCS acceleration. 
Furthermore, as the state vector-based simulations exhibit high-level of parallelism and scalability, they are widely deployed on parallel computers for accelerating the simulations. Examples include parallelizing QCS on GPUs~\cite{QuEST, cuQuantum, qHiPSTER, projectQ, Cirq}, scaling QCS across multiple machine nodes with message passing interface (MPI)~\cite{mpiQulacs, QuEST, 45qubit, cpu_gpu_communication}, scaling QCS with a larger size of quantum bits using solid-state drives (SSDs)~\cite{park2022snuqs}, and scaling QCS across different machine nodes with remote memory direct access technology (RDMA)~\cite{rdma_sim}. 


While the existing studies~\cite{cache_blocking,park2022snuqs} takes the data locality concept into account to improve the performance of QCS with extensive memory operations, these studies do not explicitly establish the ties between a processor cache and a quantum circuit for cache-aware QCS. Furthermore, they overlook the impact of the memory hierarchy when performing QCS on a multiple GPU platform. In such a case, it is hard to fully unleash the computing power of the multi-GPU platform. 

This work proposes a novel QCS toolkit for a multiple GPU platform with the distributed memory architecture. The key components of the proposed toolkit are illustrated in \figurename~\ref{fig:toolkit}. Given an input quantum circuit program representing a target quantum application, an \emph{all-in-one} (AIO) optimization module helps preprocess the input program into the format that facilitates the subsequent QCS, and an \emph{all-in-cache} (AIC) simulation module determines a suitable mechanism to facilitate QCS by managing the quantum states (maintained as state vector) on the memory hierarchy of the multi-GPU platform, including memories on different GPUs, memory regions on a GPU, and cache blocks on a GPU. 

The key idea of this work is to wisely reuse those quantum states that are placed in the memory hierarchy while updating the effects of the simulated quantum gates to take advantage of data locality. This is achieved by organizing to-be-simulated quantum gates into groups, referred to as \emph{gate blocks}. These blocks are identified by AIO based on the cache specification of the GPU for QCS, e.g., the L1 cache size. These blocks, and more importantly, their corresponding quantum states, are accessed in a cache-aware manner orchestrated by AIC in a multithreaded simulation environment. This cache-aware data orchestration is called the cache-aware simulation scheme, as will be introduced in Section~\ref{sec:workflow}. 
When the QCS is performed on a single GPU, AIC improves the simulation efficiency by increasing the data locality, where shared memory of the GPU is adopted to control the use of the on-chip GPU memory carefully. An in-memory swapping technique is adopted by AIC to place the data properly for each simulation thread. When the QCS is performed on multiple GPUs, a cross-rank swapping technique is further used to minimize the communication overhead while exchanging data among GPUs.

\begin{figure}[tb!]
\centerline{\includegraphics[width=.87\columnwidth]{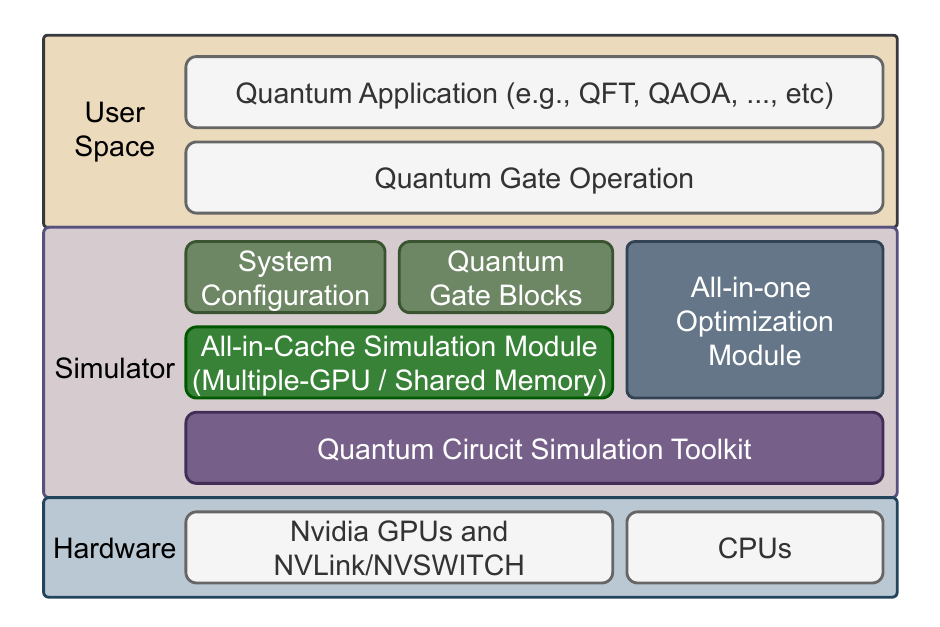}}
\caption{The quantum circuit simulation toolkit.}
\label{fig:toolkit}
\end{figure}


Our proposed toolkit has been developed and evaluated with various benchmark programs, such as micro-benchmarking on quantum gates under different qubit sizes and practical quantum applications. The performance evaluation is done on a multiple GPU server, DGX-A100 \cite{DGXA100}, with eight NVIDIA A100 \cite{A100} GPUs. Our experimental results show a significant performance improvement over the prior work, cuQuantum \cite{cuQuantum}, on the Qubit benchmark \cite{qibojit-benchmarks} and the strong scaling experiment. 
The Roofline model analysis reveals our cache-aware simulation scheme is able to shift the runtime behavior of the QCS from a memory-bound workload into a computation-bound workload. Further performance enhancement is achieved by introducing more computation resources.
The contributions of this work are summarized as follows.

\begin{itemize}
    \item A cache-aware simulation toolkit is proposed to accelerate state vector-based QCS. To the best of our knowledge, we are not aware of any other work that takes a holistic view of cache-aware QCS and cooperating different memories in the memory hierarchy of the multi-GPU architecture for high scalability to facilitate QCS.
    
    \item An all-in-one optimization module is developed to search the groups of quantum gates by taking the GPU cache organization into account. These identified gate blocks are the basic units for performing quantum circuit optimizations, such as gate fusion, to reduce the number of quantum gates. In addition, these gate blocks are basic execution units for QCS. During the simulation, the effects of the simulated quantum gates are updated on the same memory region buffered in the GPU cache, minimizing memory access overhead. 

    \item An all-in-cache simulation module contains the important mechanisms to improve the simulation performance. It is achieved by improving the data locality of each simulation thread and keeping balanced loads for multiple GPU threads (on a GPU streaming processor). In addition, a quantum-gate-dependency-aware memory swapping mechanism is developed to move data between cache and memory across streaming processors of the GPU).
    Moreover, a cross-rank data swapping mechanism is proposed to facilitate the data exchange by taking advantage of typical all-to-all communications.

    \item A series of significant performance enhancements have been achieved comprehensively compared to the state-of-the-art simulators, QuEST \cite{QuEST}, Aer \cite{Aer_sim}, and cuQuantum \cite{cuQuantum}.
    Typically, the results of the quantum applications demonstrate up to an average of 9x speedups on an NVIDIA DGX A100.

\end{itemize}

\section{Background and Related Work}
\label{sec:background}
In this section, we introduce the basic quantum gates used in QCS, along with the mechanisms of qubit swapping to improve efficiency. Subsequently, we discuss related works concerning various optimizations for QCS.

\subsection{Qubit Representation and Quantum Gates}
\label{sec:quantum_gate}
A qubit is the fundamental unit of computation in quantum computing, characterized by the state $\alpha\ket{0} + \beta\ket{1}$, where $\ket{0}=[1,0]^T$ and $\ket{1}=[0,1]^T$ in Dirac notation.
The non-deterministic behaviour of a qubit, known as superposition, allows it to exist in both states $\ket{0}$ and $\ket{1}$ simultaneously with probabilities $|\alpha|^2$ and $|\beta|^2$, respectively.
It is imperative to ensure that the probabilities must satisfy the normalization condition $|\alpha|^2 + |\beta|^2 = 1$ for a single-qubit system.

When extending this concept to multiple qubits, the quantum state becomes a superposition of $2^\mathit{N}$ basis state, ranging from $\ket{0}$ to $\ket{1}$. The $\mathit{N}$-qubit quantum state can be formulated as Equation~\ref{eq:state}, where $a_i^2$ describes the probability of the $\ket{i}$.

\begin{equation}
\ket{\phi} = \sum_{i \in [0, 2^N)} a_i \ket{i}
\label{eq:state}
\end{equation}

Quantum circuits comprise various quantum logic gates, which are the basic unit of quantum operation. They can be typically categorized into two main types: one-qubit gates and multi-qubit gates.
Consider a one-qubit gate, the matrix representation for the state vector is provided by Equation~\ref{eq:one_qubit_gate}, where the 2x2 matrix $G$ represents a one-qubit gate operating on a specific qubit and the $0_{i}$ and $1_{i}$ denote the amplitudes that the $i$-th bit in bitstring is 0 or 1.

\begin{equation}
\begin{bmatrix}
a^{\prime}_{*...*0_{i}*...*} \\
a^{\prime}_{*...*1_{i}*...*}
\end{bmatrix}
\mapsto G
\begin{bmatrix}
a^{}_{*...*0_{i}*...*} \\
a^{}_{*...*1_{i}*...*}
\end{bmatrix}
\label{eq:one_qubit_gate}
\end{equation}

The simulation of an $N$-qubit gate, represented by a $2^N \times 2^N$ matrix, involves a similar process.
For instance, consider a 2-qubit CNOT gate acting on the $i$-th and $j$-th qubits as a typical example.
The processes can be expressed as Equation~\ref{eq:cnot}.

\begin{equation}
\begin{bmatrix}
a^{\prime}_{*...*0_{i}*..*0_{j}*...*} \\
a^{\prime}_{*...*0_{i}*..*1_{j}*...*} \\
a^{\prime}_{*...*1_{i}*..*0_{j}*...*} \\
a^{\prime}_{*...*1_{i}*..*1_{j}*...*}
\end{bmatrix}
\mapsto
\begin{bmatrix}
1 & 0 & 0 & 0 \\
0 & 1 & 0 & 0 \\
0 & 0 & 0 & 1 \\
0 & 0 & 1 & 0
\end{bmatrix}
\begin{bmatrix}
a_{*...*0_{i}*..*0_{j}*...*} \\
a_{*...*0_{i}*..*1_{j}*...*}  \\
a_{*...*1_{i}*..*0_{j}*...*} \\
a_{*...*1_{i}*..*1_{j}*...*}
\end{bmatrix}
\label{eq:cnot}
\end{equation}

\subsection{Qubit Swapping and Qubit Permutation}
\label{sec:qubit_swapping}
In an $\mathit{N}$-qubit simulation, all $2^\mathit{N}$ states are constructed by combining the individual states of each qubit through a tensor product.
When qubit order is swapped, it solely impacts the arrangement of qubits within a state vector.
Equation~\ref{eq:swap_gate} expresses the operation of swapping $i$-th qubit with $j$-th qubit.
Based on the results, the state of the $i$-th qubit in state vector $a$ is identical to $j$-th qubit in statevector $a^{\prime}$.

\begin{equation}
\begin{bmatrix}
a^{\prime}_{*...*1_{i}*...*0_{j}*...*} \\
a^{\prime}_{*...*0_{i}*...*1_{j}*...*}
\end{bmatrix}
\mapsto
\begin{bmatrix}
a^{}_{*...*0_{i}*...*1_{j}*...*} \\
a^{}_{*...*1_{i}*...*0_{j}*...*}
\end{bmatrix}
\label{eq:swap_gate}
\end{equation}

The qubit swapping directly alters the qubit permutation within the provided quantum circuit, resulting in significantly different simulation performance.
For example, considering a gate permutation $(q_4q_3q_2q_1q_0)$ and applying a swap for $q_3$ and $q_2$, the permutation becomes $(q_4q_2q_3q_1q_0)$.
In practice, the simulator must swap each pair of amplitude satisfying $\ket{* 0_3 1_2 * *}$ and $\ket{* 1_3 0_2 * *}$ of the original state vector.

The deliberate integration of supplementary swaps within the circuit indicates the requirement to adjust the indexing of subsequent gates.
Although this approach needs extra operations, the qubits for the subsequent gates have already been rearranged to the least significant bits for the classical computers.
This signifies that the state represented by that qubit has been positioned closer to the computational unit, resulting in substantial improvement in data access.

\subsection{Related Work}
\label{sec:related_work}
Quantum circuit simulation can be typically categorized into state vector-based and tensor network-based approaches.
While tensor networks are limited in terms of interpretability and scalability, state vector-based QCS is widely preferred for its superior ability to represent complex quantum circuits with high scalability accurately.
Additionally, state vector-based QCS is highly parallelizable, making it ideal for execution on supercomputers~\cite{QuEST, 45qubit, qHiPSTER, cuQuantum, cache_blocking}.
To achieve effective QCS, it is essential to consider complementary optimization techniques, including \emph{maximizing parallelism}, \emph{minimizing gate operations}, and \emph{optimizing data locality}.

\emph{\textbf{Maximizing Parallelism.}}
A multitude of studies have focused on applying parallelization to enhance the simulation performance across both homogeneous and heterogeneous systems.
Within CPU architectures, multithreading and AVX vectorization techniques are implemented to exploit computational resources fully~\cite{qHiPSTER}.
In GPUs, renowned for their extensive thread capabilities, maximizing parallelism strategies leads to superior performance against homogeneous systems~\cite{QuEST, Aer_sim, cuQuantum}.
These techniques can be extended to encompass multi-rank scenarios for further performance improvement~\cite{QuEST, mpiQulacs, cpu_gpu_communication}.

\emph{\textbf{Minimizing Gate Operations.}} This technique, commonly referred to as gate fusion, is employed to merge multiple gates into a single generic gate~\cite{45qubit, qHiPSTER, cpu_gpu_communication}.
Although reducing the total number of gates can potentially decrease memory access due to the memory-bound property, these fused gates typically are less efficient and may require additional performance evaluation.
For instance, fusing an excessive number of CNOT gates in Quantum Fourier Transform (QFT) with distinct target qubits can lead to fused gates with too many qubits and exponentially growing unitary complexity.
In addition, the \emph{k-qubit} strategy~\cite{45qubit} claims high performance and scalability in specific circuits in a supercomputing cluster, yet its performance across the majority of circuits does not surpass the foundational fusion of IBM-Qiskit.
Among numerous fusion techniques, diagonal optimization is a proven technique that can simultaneously reduce the number of gates while maintaining efficiency for high performance computing~\cite{qaoa_arxiv}.

\emph{\textbf{Optimizing Data Locality.}}
To clarify the practical utility of data locality, we classified previous QCS into two types: standard simulation and all-in-rank simulation, as shown in~\figurename{~\ref{fig:simulation_type}(a) and (b), respectively.
In the standard simulation~\cite{QuEST}, the sub-state vector is duplicated across ranks for each cross-rank operation, which is considered the prototyping stage. 
In all-in-rank simulation~\cite{cuQuantum, Aer_sim, mpiQulacs}, cross-rank operations are replaced with cross-rank swaps through the qubit reordering~\cite{cache_blocking, cpu_gpu_communication} to reduce redundant data transfer.
Unfortunately, memory-intensive operations still pose a significant challenge to QCS performance.
The all-in-cache simulation, as depicted in~\figurename{~\ref{fig:simulation_type}(c), has not been entirely implemented in GPU\footnote{The qubit mapping strategy~\cite{45qubit} can enhance the arithmetic intensity by cleverly caching the workload for CPU threads, but it may fail due to the lack of consideration for the substantially massive threading characteristics of GPUs, resulting in the ongoing project remaining incomplete to this day.
}
due to the immaturity of qubit reordering for in-memory swaps and the lack of high-performance skills for in-cache operations.

\begin{figure}[tb!]
\centerline{\includegraphics[width=.98\columnwidth]{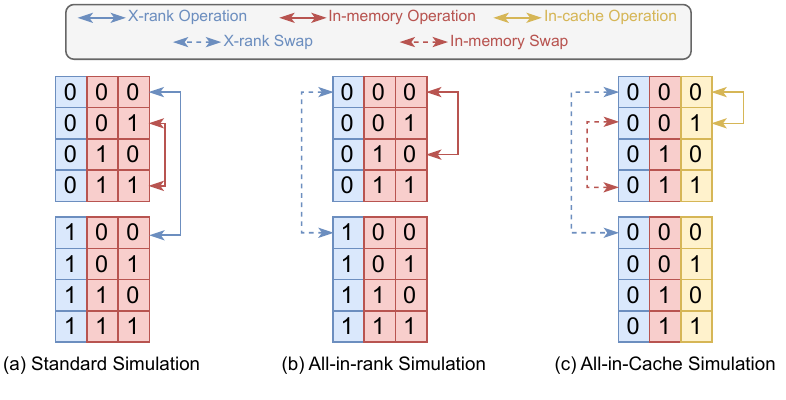}}
\caption{The types of state vector-based simulation.}
\label{fig:simulation_type}
\end{figure}

In this work, we propose a comprehensive toolkit to encompass all these three key features.
Section~\ref{sec:workflow} introduces the preliminary setup and overall simulation workflow. The AIO optimization module efficiently solves qubit reordering and gate fusion through its polynomial-time algorithm in Section~\ref{sec:aio_optization_module}.
Subsequently, Section~\ref{sec:simulation_module} presents the AIC simulation, demonstrating a series of adept implementation techniques.

\begin{figure}[btp!]
\centerline{\includegraphics[width=.9\columnwidth]{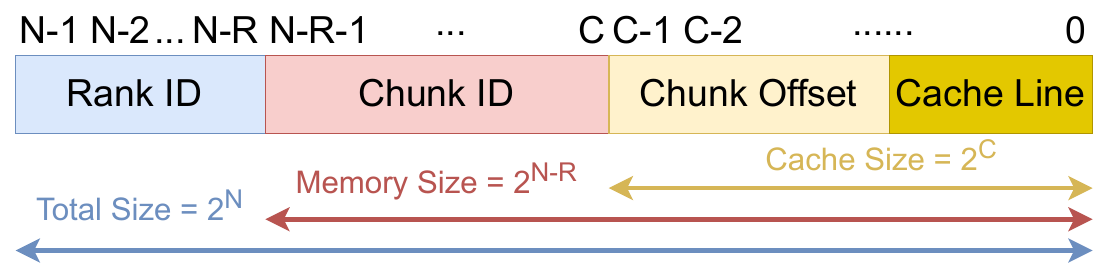}}
\caption{The partition for qubit representation.}
\label{fig:qubit_representation}
\end{figure}

\section{Preliminary Setup and Simulation Workflow}
\label{sec:workflow}

The development of the simulation toolkit with robust performance and scalability relies on a well-suited partitioning strategy as the preliminary setup and a smooth workflow.

For preliminary setup, considering a state vector of size $2^\mathit{N}$ as illustrated in \figurename{~\ref{fig:qubit_representation}}, the entire state is initially partitioned into $2^\mathit{R}$ sub-states for the multi-rank setup.
The $\mathit{R}$ preceding index bits determine the rank ID, ensuring that $2^{\mathit{N}}$ amplitudes are uniformly distributed across $\mathit{R}$ ranks without overlaps.
Within each rank, the $2^{\mathit{N-R}}$ state is further separated into smaller sub-states, each containing $2^\mathit{C}$ amplitudes, serving as the basic unit of our simulation. 
Additionally, the remaining $\mathit{CL}$ bits indicate cache line bits, with amplitudes differing in cache line bits residing in distinct cache lines, which will be further discussed in Section~\ref{sec:sqs}.
With this setup, we gain the necessary insights to allocate different target qubits of gates, denoted as \emph{targ}, to distinct operations and swaps in \figurename{~\ref{fig:simulation_type}}(c).

\emph{\textbf{In-cache Operation}}. When $\emph{targ}<\mathit{C}$, it indicates that the indexing pairs for the $\emph{targ}$ in the state vector can be located within the cache size.
To increase data reuse, the simulation scheme is designed to load data into the cache size and perform consecutive computations until this condition is no longer met.
For example, consider a $3$-qubit simulation system with $2^{1}$ cache size ($\mathit{C}$=1), where $\emph{targ}$ for the current 1-qubit gate is zero.
In this scenario, the `000' state and its corresponding `001' state are considered an indexing pair, and they both need to be read into the computing unit for calculation.
Then, if all $\emph{targ}$s of subsequent gates are also zero, in-cache operations can perform consecutive gate operations without writing back to memory, unlike other SOTA simulators~\cite{QuEST, cuQuantum, qHiPSTER, projectQ, Cirq}.

\emph{\textbf{In-memory and Cross-rank Swap.}} On the contrary, when $\emph{targ}\geq\mathit{C}$, the paired sub-states fall outside the cache size.
Instead of simulating the gate operations, we employ swapping techniques to exchange the sub-states from local memory and remote ranks to the cache region, keeping a high level of data locality for our in-cache operation.
Moreover, since the previous swapping implementations~\cite{Cirq, cuQuantum, mpiQulacs} fail to meet our high performance and scalability requirements, we redesign the in-memory and cross-rank swaps and elucidate their implementation in Section~\ref{sec:sqs} and Section~\ref{sec:xdev_swapping}, respectively.

\emph{\textbf{Gate Block Finding Algorithm}}.
A \emph{gate block} is defined as a set that gathers a sequence of gates suitable for in-cache operations.
To achieve optimal cache capability, the objective can be simplified to maximize the number of gates in each gate block, which requires a preprocessing for the given circuit.
The previous related approaches~\cite{loser, cache_blocking, cpu_gpu_communication} are overly intricate and yield inadequate results, thereby negatively impacting overall performance in a single rank.
We overhauled the gate block finding algorithm to tackle gate fusion and qubit reordering simultaneously, resolving these issues effectively.

Overall, our simulation workflow is illustrated in \figurename{~\ref{fig:workflow}}.
Initially, a raw circuit is fed into an AIO optimization module for preprocessing. 
This module is designed to organize as many nearby gates as possible into a gate block to facilitate in-cache gate operations while incorporating in-memory and cross-rank swaps between gate blocks to ensure data integrity.
Subsequently, the optimized quantum circuit is processed by the AIC simulation module, which sequentially executes the operations of each gate block, producing the simulation results.

\begin{figure}[tb!]
\centerline{\includegraphics[width=.85\columnwidth]{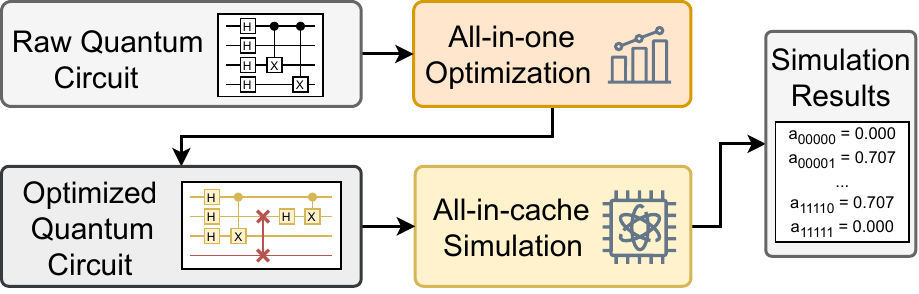}}
\caption{The workflow of our quantum circuit simulation.}
\label{fig:workflow}
\end{figure}

\section{All-in-one Optimization Module}
\label{sec:aio_optization_module}

\begin{figure*}[tb!]
\centerline{\includegraphics[width=1.95\columnwidth]{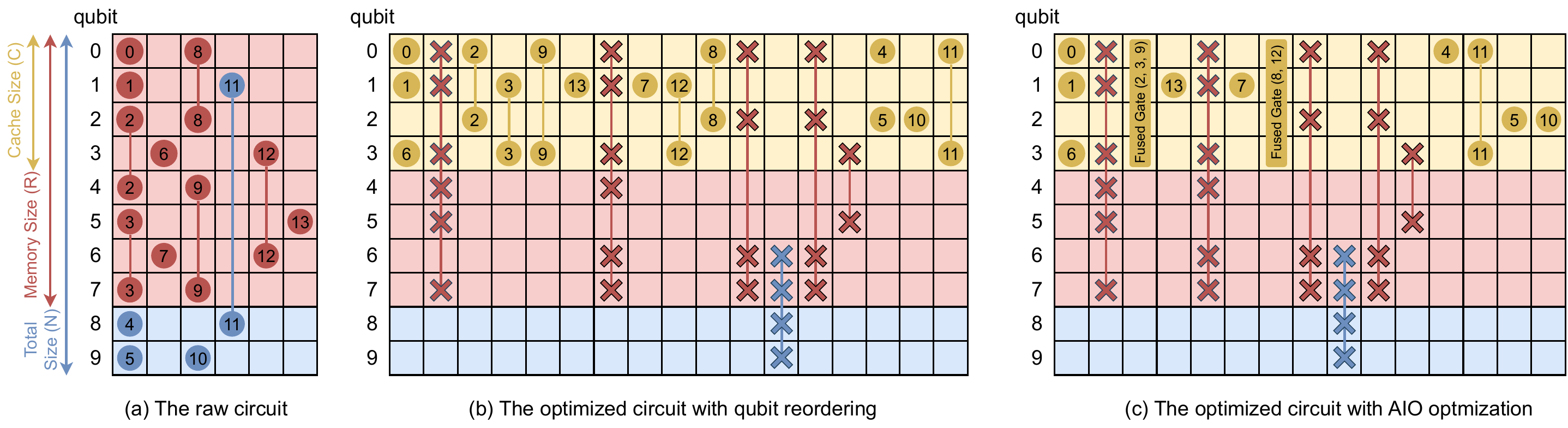}}
\caption{The circuit diagram for the simulator.}
\label{fig:swapExam}
\end{figure*}

From a broad perspective, the existing quantum circuit optimization techniques, such as gate fusion and qubit reordering, can be generalized into a unified mathematical optimization problem with minor adjustments to constraints.
That is, these methods seek to minimize a specific cost function within a given circuit.
For instance, the cost objective is to minimize the count of fused gates and the frequency of data transfers for gate fusion and qubit reordering, respectively.

To formalize this process, we define a set, denoted as gate block (\textit{GB}), to aggregate these specific gates.
The constraints typically entail dependencies ($\textit{dep}$) and the total number of involved target qubits (\textit{chunkSize}), as shown in Equation~\ref{eq:gb}.

\begin{equation}
\begin{array}{ll}
\label{eq:gb}
\textit{GB} = \{ g \in \textit{gateList} \}, \ \| \underset{g \in \textit{GB}}{\bigcup} \textit{dep}(g) \| \leq \textit{chunkSize}
\end{array}
\end{equation}

Then, as illustrated in Equation~\ref{eq:gbs}, all determined gate blocks are included in a more extensive set \textit{GBs}. where the intersection of any two gate blocks is an empty set, and the union of all gate blocks is equivalent to \textit{gateList}.

\begin{equation}
\begin{array}{ll}
\label{eq:gbs}
\textit{GBs} = \{ \textit{GB}_i \} , \ \textit{GB}_i \cap \textit{GB}_j = \varnothing, \forall i<j  \\ 
\end{array}
\end{equation}

By applying the definitions provided above, the optimization problems can be efficiently re-formulated by minimizing the number of \textit{GBs}. 
To facilitate the implementation of the optimization process programmatically, we transform it as an equivalent maximization problem, seeking to maximize the average size of \textit{GB} within \textit{GBs}, as depicted in Equation~\ref{eq:min}.

\begin{equation}
\begin{array}{lll}
\label{eq:min}
\underset{\textit{GBs}}{\text{argmin}} \ \|\textit{GBs}\| \iff \underset{\textit{GBs}}{\text{argmax}} \ \sum\limits_{\textit{GB} \in \textit{GBs}} \frac{  \|\textit{GB}\| } { \|\textit{GBs}\| } \\
\end{array}
\end{equation}

The pseudo-code presented in Algorithm~\ref{algo:find_gb} demonstrates how the finding process fits the maximization problem within polynomial time.
During each iteration (in Line 4), the algorithm updates dependencies in Line 6 and determines the maximum \textit{GB} in Line 7.
Specific functionality for different problems is dealt with in Lines 8-13.
Following the completion of the algorithm, the optimized list \textit{GBs} is returned in Line 15.

\begin{algorithm}[hbt!]
  \caption{Pseudo-code to find the gate blocks.}
  \label{algo:find_gb}
  \SetKwInOut{kwInput}{Input}
  \SetKwInOut{kwOutput}{Output}
  \SetKwProg{Fn}{Function}{:}{end}
  \SetKw{Continue}{continue}

  \Fn
    {findGBs({\textit{gateList}, $\mathit{N}$, \textit{chunkSize}, isFusion}) } {
      chunkSet $\gets$ $\{q_{0}, \dots, q_{\textit{chunkSize}-1}\}$

      \textit{GBs}.append(setupGB(\textit{gateList}, chunkSet))

      \While{\textit{gateList} $\neq$ None} {
        prevSet $\gets$ chunkSet

        updateDependency(\textit{gateList}, $\mathit{N}$)

        chunkSet $\gets$ findMaxGate(\textit{gateList}, \textit{chunkSize})

        \If {isFusion} {
            doFusion(\textit{GBs}, $\mathit{N}$, \textit{chunkSize}, chunkSet)

            \Continue
        }

        insertQubitSwaps(\textit{GBs}, prevSet, chunkSet)

        \textit{GBs}.append(setupGB(\textit{gateList}, chunkSet))

      }

      \Return \textit{GBs}
    }

\end{algorithm}

Algorithm~\ref{algo:aio_opt} outlines the pseudo-code for the complete all-in-one optimization, utilizing $\mathit{R}$, $\mathit{C}$, $\mathit{F}$ to denote the relationship of memory size, cache size, and the scope of gate fusion, respectively.
This optimization consists of three-level progress, each employing the \textit{findGBs}() function with decreasing order, sequentially occurring in Lines 1, 4, and 8.
With each progression, the search scope becomes more refined, ultimately leading to an optimization duration that ordinarily occupies less than 1\% of the total simulation time.

In~\figurename{~\ref{fig:swapExam}}(a), there is a raw circuit made up of fourteen quantum gates.
Each symbol in the circuit corresponds to a different gate with a unique ID number, and the position of the symbol represents the targeted qubit.
Simulating this raw circuit with the previous simulators requires ten memory-level accesses and four cross-rank-level accesses. 
Even with cache-level optimization, the circuit only engages four cache-level accesses, indicating an inefficient exploitation of cache characteristics.

\figurename{~\ref{fig:swapExam}} (b) demonstrates the optimized circuit resulting from qubit reordering by our proposed swapping operations (denoted by crosses).
All quantum gates in the circuit can be accommodated within the cache region, requiring five memory-level accesses and only one cross-rank-level access for swapping operation.
Furthermore, upon enabling all optimization features, as shown in~\figurename{~\ref{fig:swapExam}} (c), gates can be seamlessly fused within the cache region with optimal circuit depth.
This AIO module can guarantee that computations predominantly maximize the cache utilization while minimizing the number of gate operations with high scalability.


\begin{algorithm}[hbt!]
  \caption{Pseudo-code of all-in-one optimization.}
  \label{algo:aio_opt}
  \SetKwInOut{kwInput}{Input}
  \SetKwInOut{kwOutput}{Output}

  \kwInput{\textit{gateList}, $\mathit{N}$, $\mathit{R}$, $\mathit{C}$, $\mathit{F}$, \textit{isFusion}}
  \kwOutput{\textit{GB}}

\textit{devGBs} $\gets$ \textit{findGBs}(\textit{gateList}, $\mathit{N}$, $\mathit{N}$-$\mathit{R}$, 0)\tcp*{1-level}

  \For(\tcp*[f]{2-level}){devGB $\in$ devGBs} {
    \textit{gateList} $\gets$ setupList(\textit{devGB})

    \textit{GB} $\gets$ \textit{findGBs}(\textit{gateList}, $\mathit{N}$, $\mathit{C}$, 0) 

    \If {isFusion} {
      \For(\tcp*[f]{3-level}){GB $\in$ GBs} {
        \textit{gateList} $\gets$ setupList(\textit{GB})

        \textit{fusedGBs} $\gets$ \textit{findGBs}(\textit{gateList}, $\mathit{N}$, $\mathit{F}$, 1) 
      }
      \textit{GBs} $\gets$ \textit{fusedGBs}
    }
  }

  \SetKwFunction{SFA} {}
  \SetKwProg{Fn}{Function}{:}{end}
  \SetKw{Continue}{continue}











\end{algorithm}


\section{All-in-Cache Simulation Module}
\label{sec:simulation_module}

The pseudo-code for the AIC simulation process is presented in Algorithm~\ref{algo:simulation_flow}.
This simulation takes an optimized circuit as the input and iterates through the gate blocks, as indicated in Line 1.
When encountering different types of gate blocks, the simulation performs cross-rank swaps (Line 3), in-memory swaps (Line 5), and in-cache operations with the block-by-block scheme (Lines 7-11).
In the pursuit of optimal performance on NVIDIA multi-GPUs, all implementations are tailored for CUDA, integrating a comprehensive range of hardware and software techniques in the subsequent subsection.



\begin{algorithm}[hbt!]
  \caption{Pseudo-code of all-in-cache simulation.}
  \label{algo:simulation_flow}
  \SetKwInOut{kwInput}{Input}
  \SetKwInOut{kwOutput}{Output}
  \kwInput{The initialized stateVector and gateBlocks}
  \kwOutput{The simulated amplitude in stateVector}
  \For(\tcp*[f]{Scan gate block}){\textnormal{gateBlock $\in$ gateBlocks}} {
    \uIf {\textnormal{isCrossRankSwap(gateBlock)}} {
        stateVector$\gets$doCrossRankSwap(stateVector)
    }
    \uElseIf {\textnormal{isInMemSwap(gateBlock)}} {
        stateVector $\gets$ doInMemSwap(stateVector)
    }
    \Else (\tcp*[f]{Do quantum gate operations in cache}){
        \For {\textnormal{chunk $\in$ stateVector}} {
            \For {\textnormal{gate $\in$ gateBlock}} { 
                chunk $\gets$ doOperation(gate, chunk)
            }
        }
    }
  }  
\end{algorithm}


\subsection{In-cache Gate Operations in Block-by-block Scheme}
\label{sec:gate_operations}
In general, quantum gates are the foundational unit of computational operations in quantum computing and shape the landscape of quantum algorithms.
Within these algorithms, a quantum circuit is formed by applying diverse types of quantum gate operations to tackle practical challenges in the specific application domains.
The optimization of gate operations can lead to heightened efficiency, enabling simulations to complete complex tasks in less time.

In previous quantum circuit simulators, gate operations are computed using matrix multiplications and simulated in a gate-by-gate scheme.
However, those legacy approaches fail to utilize the computational mechanisms of classical computers fully, thereby impeding the adoption of advanced high-performance techniques.
To rectify the situation, our simulator introduces a series of strategies to expedite simulation and eliminate performance bottlenecks, delineated as follows.

\emph{\textbf{Increase data locality.}}
To enhance data reuse, the state vector is partitioned into non-overlapping chunks, where each GPU thread block exclusively processes states within its designated chunk.
Gates are grouped into gate blocks through our AIO optimization module, which guarantees that all operations within a block can be applied to a chunk without requiring data exchange beyond the chunk boundaries.
The simulation follows the determined gate block sequentially, resulting in a substantial improvement in data locality compared to alternative simulators.
Moreover, we actively place data into shared memory to explicitly control the use of the on-chip GPU memory.
This approach enables improved management of GPU performance tuning and thread block allocation to stream multiprocessors, leading to enhanced overall performance.

\emph{\textbf{Decrease stall barriers.}} 
In open-source GPU simulators~\cite{QuEST, cpu_gpu_communication, mpiQulacs}, each thread typically manages a single quantum state in its implementation.
However, this approach can lead to uneven task distribution when operations involve varying numbers of target and control qubits. 
As a result, most threads in a thread block must wait at barriers without performing any computations to preserve data integrity, which negatively impacts performance.
To overcome this challenge, we adopt loop unrolling, allowing each thread to handle multiple states to utilize computational resources fully.
This technique is applied to each gate individually, ensuring the optimal unrolling factors for minimizing thread management and loop overhead without unnecessary stall barriers.

\subsection{In-memory Swapping (IMS)}
\label{sec:sqs}
In our cache-aware implementation of simulation, the predominant factor driving overall performance improvement is the qubit swapping between the cache and memory region.
While the most naive approach involves employing either the native swap gate or the fused swap gate, these approaches necessitate an excessive number of swap gates or the closure of numerous threads to guarantee correct simulation results.
Consequently, the simulation time in a single rank is even worse than when these approaches are not used.

In-memory Swapping (IMS) is proposed as a critical solution to swap all required qubits once and for all.
To efficiently implement this swapping operation in multi-threaded architectures, it is not sufficient to intuitively apply threads to locate the corresponding positions for data exchange due to a multitude of scattered memory accesses.
Contrarily, our IMS is designed to thoroughly utilize the characteristics of cache lines, which serve as the basic unit for transferring data between cache and memory, as detailed in Algorithm~\ref{algo:sqs}.

The $2^N$ state vector can be parallelized by $2^N$ GPU threads, with a variable $t$ representing each thread ID in Line 1.
Then, the pair of cache-friendly indexes $m$ and $n$ can be calculated by~\emph{bitshift()} and~\emph{bitswap()} functions in Lines 2-3.
The \emph{bitshift()} function facilitates the shifting of the required swapped-out bit $c$ from the left of the most significant bit of $\mathit{CL}$ to the positions that match the lowest $c$ qubits being swapped in.
The \emph{bitswap()} function is the original mapping rule of quantum computing for the qubit swapping.
After this stage, \textit{stateVector} is swapped with these indexes in Lines 4-6.

\figurename{~\ref{fig:sqs}} illustrates an example of the IMS process with the swapping out set \{${q_0, q_2, q_3, q_4, q_5}$\} and swapping in set \{${q_6, q_7, q_9, q_{10}, q_{11}}$\}.
In this case, the $t_3$ and $t_4$ of each thread correspond to the $q_6$ and $q_7$ of the original qubit representation.
Consequently, the swapping rule is used to map them to the swapped-out indexes $q_0$ and $q_2$.
All bits of $m$ and $n$ that are in $\mathit{CL}$ must be mapped from the bits of thread block offsets (e.g., The indexes $t_0, ..., t_5$ of thread marked in red within \figurename{~\ref{fig:sqs}}), providing clear evidence for this optimization.

It is worth noting that our approaches can seamlessly transition to homogeneous systems by fine-tuning the value $\mathit{CL}$ and parallelism parameters to match specific CPU architectures.

\begin{figure}[tb!]
\centerline{\includegraphics[width=.85\columnwidth]{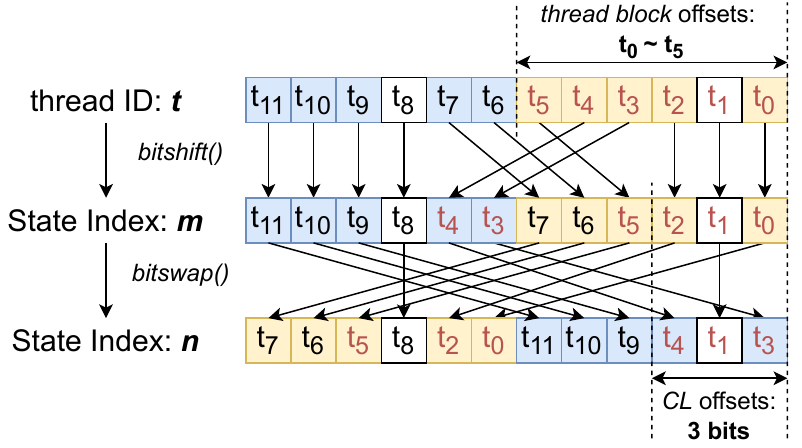}}
\caption{An example of the in-memory swapping process.
}
\label{fig:sqs}
\end{figure}

\begin{algorithm}
    \caption{Pseudo-code of in-memory swappipng}
    \label{algo:sqs}
    \SetKwInOut{KwInput}{Input}
    \SetKwInOut{KwOutput}{Output}

    \KwInput{The indices of qubits that are swapped in the chunk ($A$) and swapped out of the chunk ($B$).
    }

    \For{$t \gets 0$ \KwTo $2^N - 1$} {
        $m$ $\gets$ bitshift($t$, $A$, $B$, $\mathit{CL}$) \\
        $n$ $\gets$ bitswap($m$, $A$, $B$) \\
        \If {$m > n$} {
            swap($\textit{stateVector}[m]$, $\textit{stateVector}[n]$) \\
        }
    }
    
\end{algorithm}


\subsection{Cross-rank Swapping (XRS)}
\label{sec:xdev_swapping}
Cross-rank swapping (XRS) constitutes the communication technique for applying the swapped qubit to a local rank.
Compared to IMS, XRS is commonly used to expand storage space across multiple hardware ranks~\cite{QuEST, cache_blocking, cpu_gpu_communication, mpiQulacs, JUQCS, intel_QS, cuQuantum}.
As depicted in \figurename{~\ref{fig:nccl}}(a), the standard implementation is accomplished through the repetition of three phases, including gathering, cross-rank transfer, and scattering.
This approach necessitates that endpoints manage potentially fragmented data transfer with two supplementary buffers.
Consequently, it not only degrade overall operational efficacy but also compels the allocation of limited residual memory resources.


Fortunately, these challenges are effectively alleviated with the adoption of our AIO optimization. This optimization ensures that data sent to a peer is concentrated, transforming the XRS operation into an in-place all-to-all communication.
Compared to the standard implementation, this eliminates the need for send buffers and replaces time-consuming gather/scatter primitives with a single memory copy, as shown in \figurename{~\ref{fig:nccl}}(b).
Moreover, to achieve optimal cross-rank transfer performance in our multi-GPU environment, we utilize the NVIDIA Collective Communication Library (NCCL) as the underlying communication library\footnote{For portability CPUs and AMD GPUs, the API in NCCL can be seamlessly replaced with MPI and RCCL libraries, respectively.}.


The Algorithm~\ref{algo:xrs} shows the complete process of XRS swap. The process is iterative all-to-all operations
\footnote{In NCCL, there is no direct all-to-all primitive. Instead, multiple send and receive primitives are encapsulated within special group functions (ncclGroupStart/ncclGroupEnd)~\cite{nccl_p2p}. These group functions can merge these primitives to enhance performance~\cite{nccl_group}.}
in lines 2-4, followed by memory copy operations in lines 5-9. 
Upon reaching the buffer of other GPU ranks, the data are copied back to the main memory to ensure that subsequent instructions can smoothly utilize the advantages of shared memory for subsequent gate operations.
Ultimately, the implementation achieves full hardware bandwidth utilization while keeping high precision and scalability for cross node parallelism.

\begin{algorithm}
\caption{Pseudo-code of cross-rank swapping}
\label{algo:xrs}
    \SetKwInOut{KwIn}{Input}
    \SetKwInOut{KwOut}{Output}
    \KwIn{The number of swap pairs ($S$) and the receive buffer size ($2^B$).
    }
   

    
    \For{$\textit{offset} \gets 0 $ to $2^{N-R-S}$ by $2^{B-S}$}{
        \For{$grp \gets 0$ to $2^{R-S}-1$}{ 
            allToAll() \tcp{All-to-all data transfer}
        }
        \For{$grp \gets 0$ to $2^{R-S}-1$}{
            \For{$r \gets 0$ to $2^{S}-1$} {
                memcpyRecvBufToSv()
            }
        }

    }
\end{algorithm}

\begin{figure}[tb!]
\centerline{\includegraphics[width=.95\columnwidth]{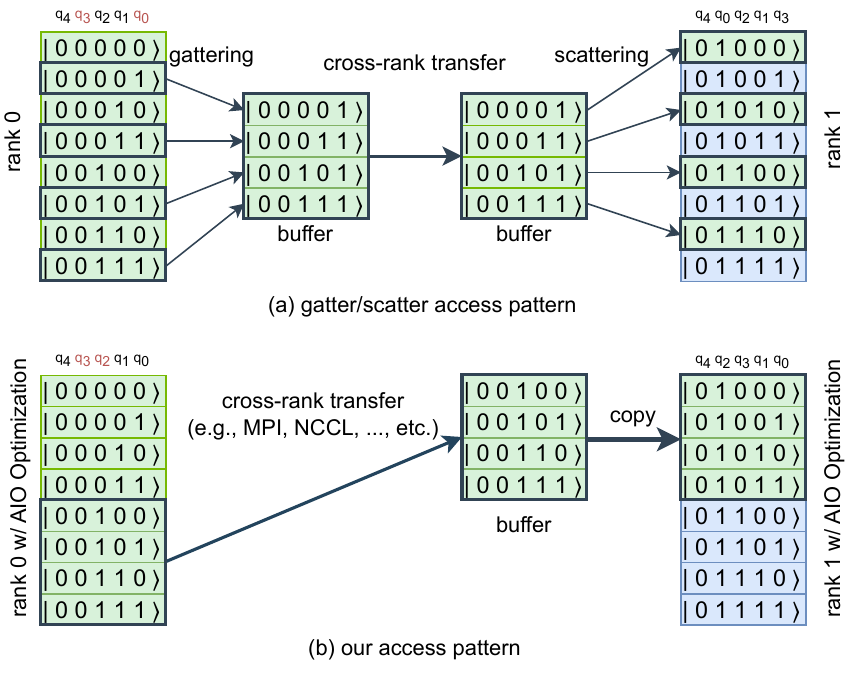}}
\caption{Data communication for cross-rank swapping.}
\label{fig:nccl}
\end{figure}

\section{Performance Evaluation}
\label{sec:evaluation}
Section~\ref{sec:env_setup} introduces the experimental environment and benchmark setup for both quantum gates and circuits.
Subsequently, Section~\ref{sec:full_opt} presents the overall performance improvements compared to the state-of-the-art work with the same inter-GPU communication link.
To prove the strength of our approach, experiments are conducted with various qubits and rank configurations in Section~\ref{sec:h_benchmark}.
Moreover, to emphasize versatility and execution efficiency for each implementation, we present benchmarks for various gates and circuits without additional transpilations in Section~\ref{sec:gate_benchmark},
Lastly, Section~\ref{sec:performance_analysis} and Section~\ref{sec:roofline_model} offer a comprehensive performance analysis of our simulator.


\begin{figure*}[htb]
\captionsetup[subfloat]{farskip=2pt,captionskip=1pt}
\centering
\subfloat[Strong scaling benchmark on 5-level QAOA.]{\includegraphics[width=0.44\textwidth]{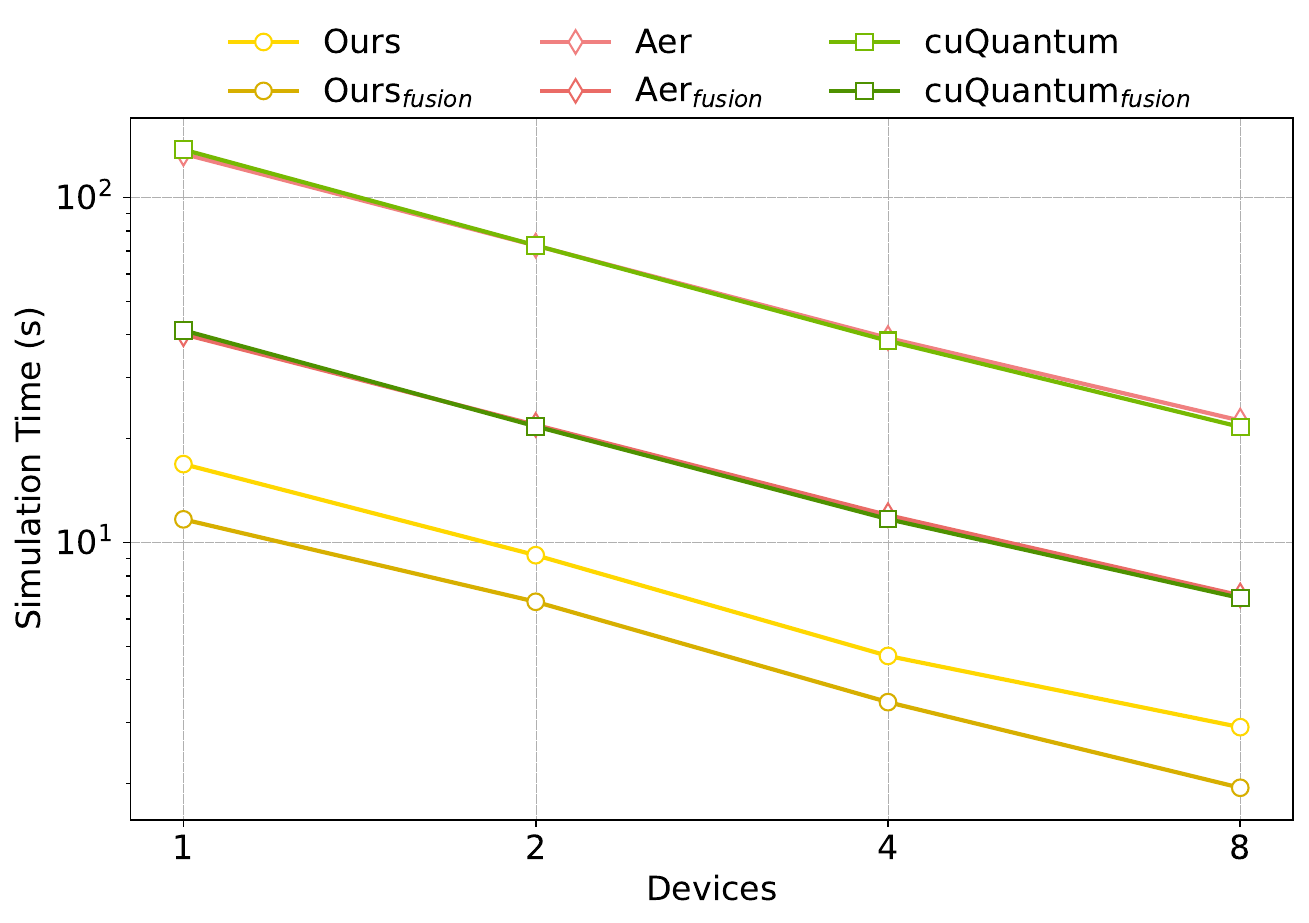}
\label{fig:qaoa_time}
}\hfill
\subfloat[Strong scaling benchmark on QFT.]{\includegraphics[width=0.44\textwidth]{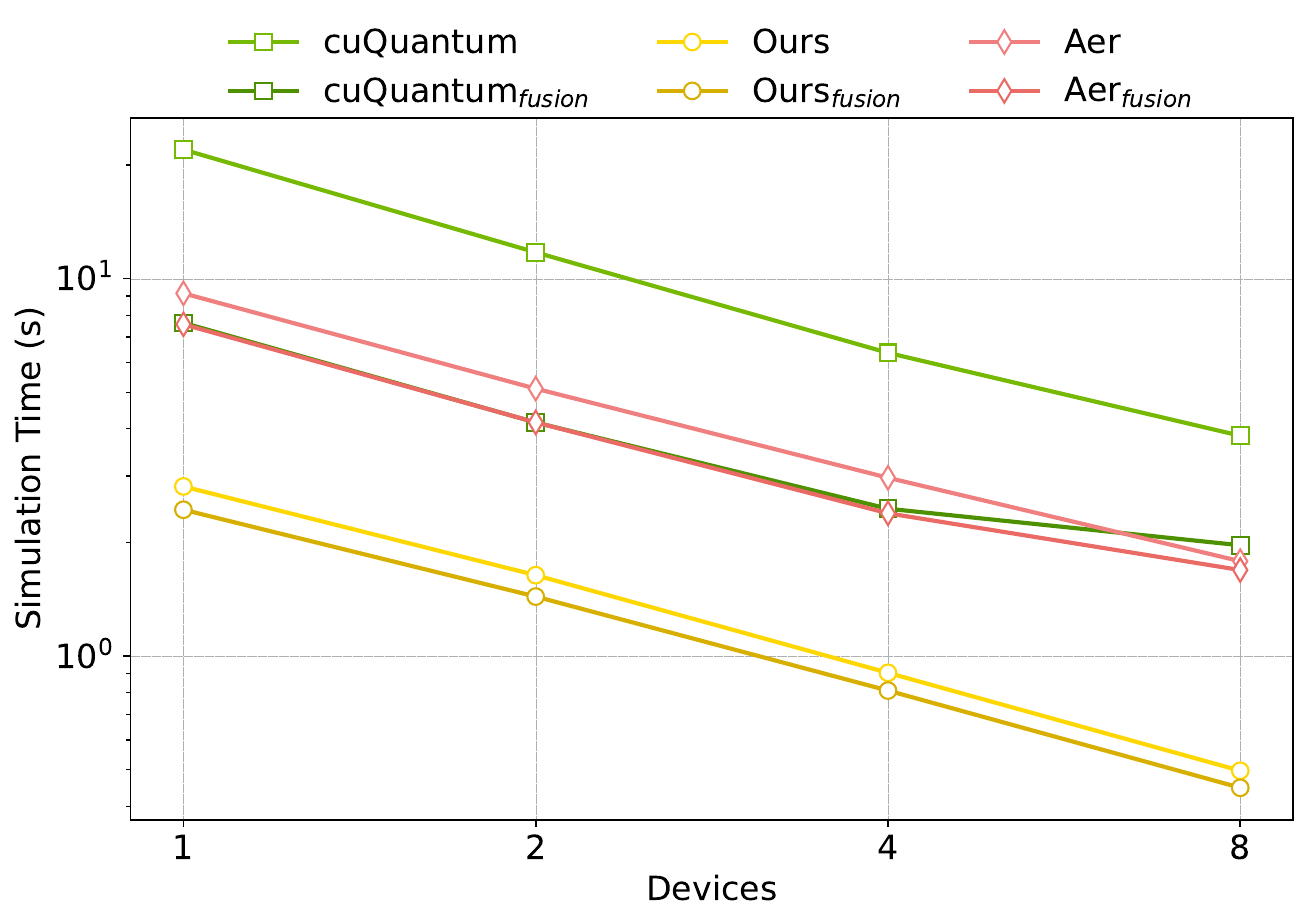}
\label{fig:qft_time}
}\\[-1ex]
\caption{
Overall simulation time on GPU A100s for QAOA and QFT with the subscript \emph{fusion} indicating gate fusion activation.
}
\label{fig:circuit_time}
\end{figure*}

\begin{figure*}[htb]
\captionsetup[subfloat]{farskip=2pt,captionskip=1pt}
\centering
\subfloat[Qubit benchmark.]{\includegraphics[width=0.44\textwidth]{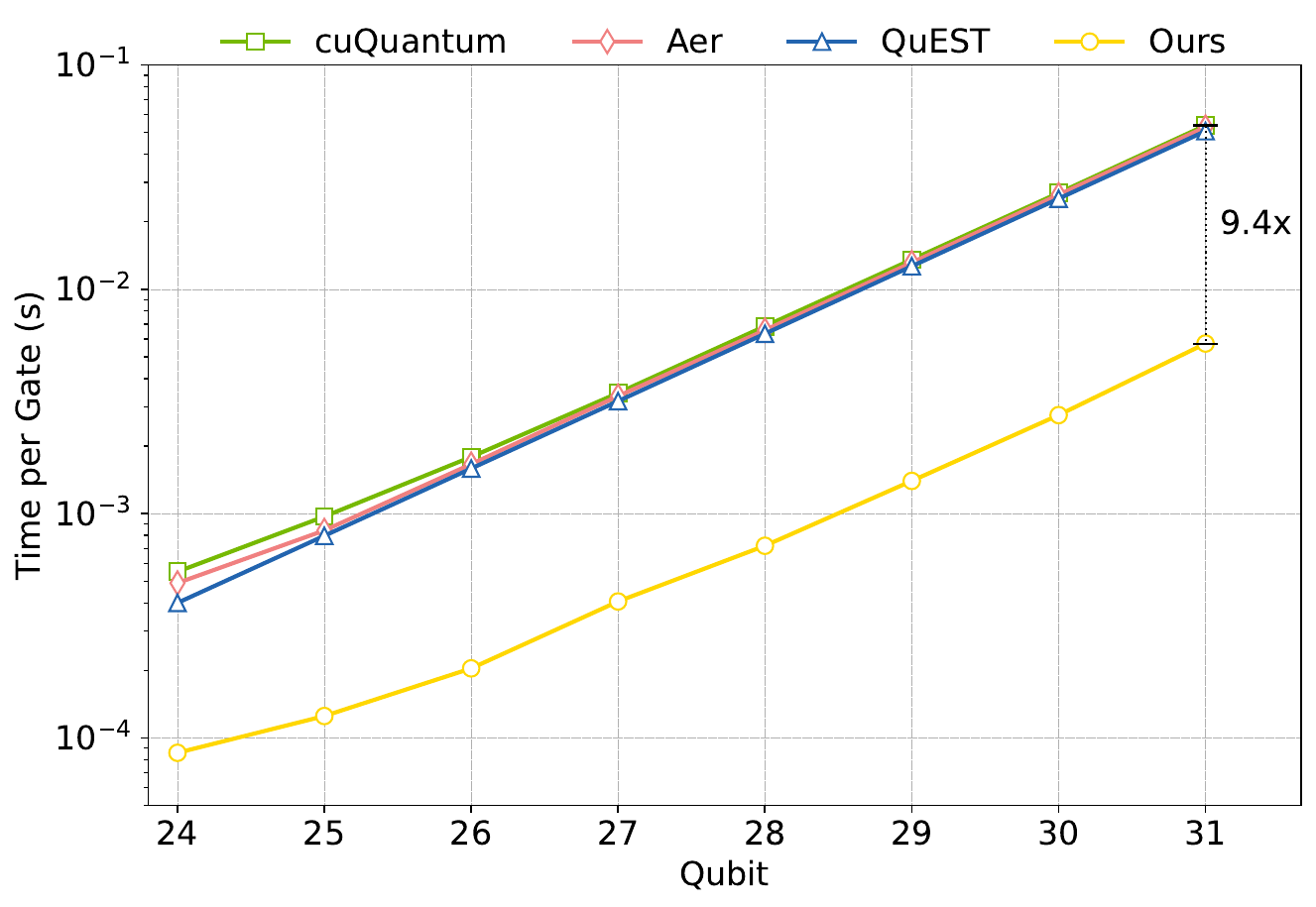}
\label{fig:qubit_benchmark}
}\hfill
\subfloat[Strong scaling.]{\includegraphics[width=0.44\textwidth]{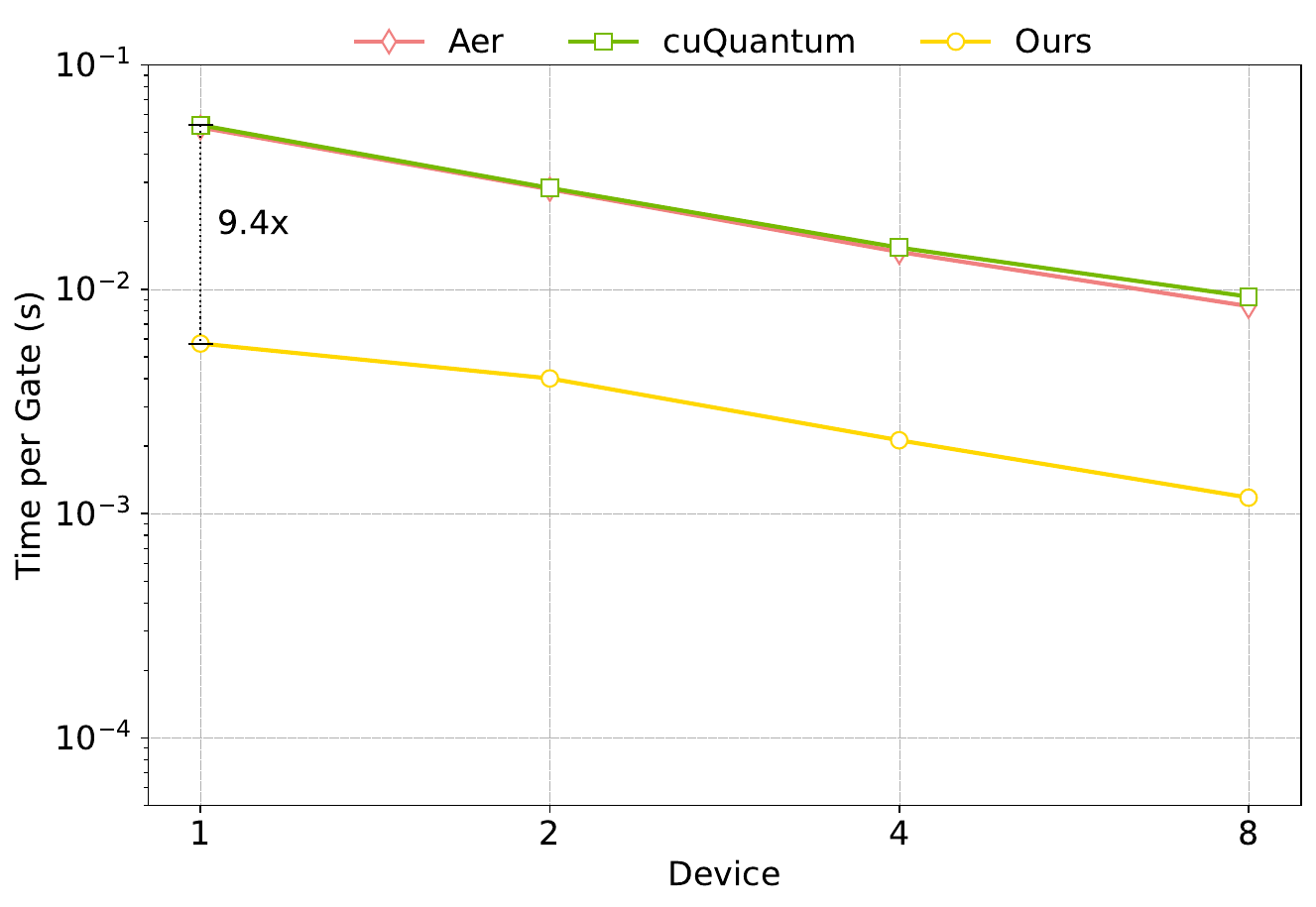}
\label{fig:strong_scale}
}\\[-1ex]
\caption{The qubit benchmark and strong scaling test for 31-qubit Hadamard gates.}
\label{fig:h_benchmark}
\end{figure*}

\subsection{Experimental Setup}
\label{sec:env_setup}
The experiments
were conducted on a DGX-A100 workstation~\cite{DGXA100} featuring two AMD EPYC 7742 CPUs with a total of 128 cores and 8 NVIDIA A100 SXM GPUs, each equipped with 40 GB of memory and interconnected via NVSwitch chips~\cite{nvlink}, delivering a bi-directional bandwidth of 600 GB/s.
The operating system is Ubuntu 22.04 with kernel version 5.15.0-1044-nvidia, alongside CUDA toolkit version 12.3.107 and NCCL library version 2.20.3~\cite{nccl}.

To ensure consistency and fairness in evaluating simulation performance, the average execution time is derived from 10 runs, adopting \emph{double}-precision floating-point numbers.
The experiments employ a 31-qubit configuration, representing the maximum qubit capacity of an A100 GPU to cover the majority of performance results.
For comparisons involving multiple ranks, we utilize the IBM Aer simulator with CUDA thrust~\cite{thrust} and cuQumatum as the backend. 
In single-rank comparisons, QuEST is included for examination.


To benchmark the quantum gate, we use the standard gates listed in Table~\ref{tab:gate_benchmark}. Each qubit undergoes the application of the 10 specified gate operations.
During performance calibration for scalability, we employ the Hadamard (H) gate, which is most frequently utilized in quantum initialization.
\begin{table}[htb]
\centering
\setlength{\tabcolsep}{0.3mm}
    \caption{The Gate Benchmarks.}
    \label{tab:gate_benchmark}
    \begin{tabular}{|c|c|c|c|}
\hline
Gates & Meanings & Gates & Meanings\\ \hline
H & Hadamard & U & Generic 1-qubit gate \\ \hline
X & Pauli-X bit flip & CX & Controlled-NOT \\ \hline
CP & Controlled phase & SWAP & 1-qubit Swap \\ \hline
RX & X-axis rotation & RY & Y-axis rotation \\ \hline
RZ & Z-axis rotation & RZZ & 2-qubit ZZ rotation \\ \hline
    \end{tabular}
\end{table}

For quantum circuits, Table~\ref{tab:circuit_benchmark} outlines them across a range of applications and problem sizes. 
These circuits were obtained using the open-source circuit generator~\cite{qibojit-benchmarks}.
For the overall performance, we target Quantum Fourier Transform (QFT) and Quantum Approximate Optimization Algorithm (QAOA) with p = 5, while utilizing up to 8 GPUs for strong scaling test.

\begin{table}[htb]
\centering
\setlength{\tabcolsep}{0.4mm}
    \caption{The Circuit Benchmarks on 31-qubit.}
    \label{tab:circuit_benchmark}
    \begin{tabular}{|c|c|c|c|}
\hline
Abbr. & Name & Domain & Gates \\ \hline

BV~\cite{bv}   & Bernstein-Vazirani & Hidden Subgroup & 92 \\ \hline
HS~\cite{hs}  & Hidden Shift Circuit & Hidden Subgroup & 149 \\ \hline
QAOA~\cite{qaoa} & \makecell{Quantum Approximate \\ Optimization Algorithm} & Optimization & 2,511 \\ \hline
QFT~\cite{qft} & \makecell{Quantum Fourier \\ Transform} & Hidden Subgroup & 496 \\ \hline
QV~\cite{qv} & Quantum Volume & Validation & 495 \\ \hline
SC~\cite{sc} & Supremacy Circuit & Optimization & 285 \\ \hline
VC~\cite{vc} & Variational Circuit & Optimization & 276 \\ \hline
    \end{tabular}
\end{table}

\subsection{Overall Performance Speedup}
\label{sec:full_opt}
We present the complete optimization of our simulator for QAOA and QFT compared with the Aer simulator and cuQuantum, as illustrated in \figurename~\ref{fig:circuit_time}.
In scenarios exclusively featuring raw circuits, both Aer and cuQuantum require over 130 seconds to simulate 31-qubit 5-level QAOA, whereas our work is completed in just 17.5 seconds, achieving a 7.5x improvement.
For the highly qubit-dependent QFT, our simulator finishes in 3.6 seconds, surpassing Aer and cuQuantum by 6.4 and 2.8 times, respectively.
Additionally, our work on a single GPU can simulate QAOA dramatically faster than Aer and cuQuantum, even when they are utilizing 8 GPUs.


Upon enabling fusion optimization, our simulator experiences further enhancement, achieving an 11.5x optimization compared to running Aer and cuQuantum without transpilation for QAOA in a single rank. With the activation of full IBM-Qiskit circuit optimization, the memory access frequency of Aer and cuQuantum is alleviated, yet our single simulator execution time remains faster than their 4-GPU setup.


In consideration of the overall trend, our simulator demonstrates both high speed and scalability.
Interestingly, our proposed block-by-block scheme, without gate fusion on a GPU, consistently exhibits superiority over others incorporating gate fusion on 2 GPUs.
These observations imply that our simulator operates at a superior level relative to previous SOTA works.






\begin{figure*}[htb]
\captionsetup[subfloat]{farskip=2pt,captionskip=1pt}
\centering
\subfloat[Gate benchmark.]
{\includegraphics[width=0.43\textwidth]{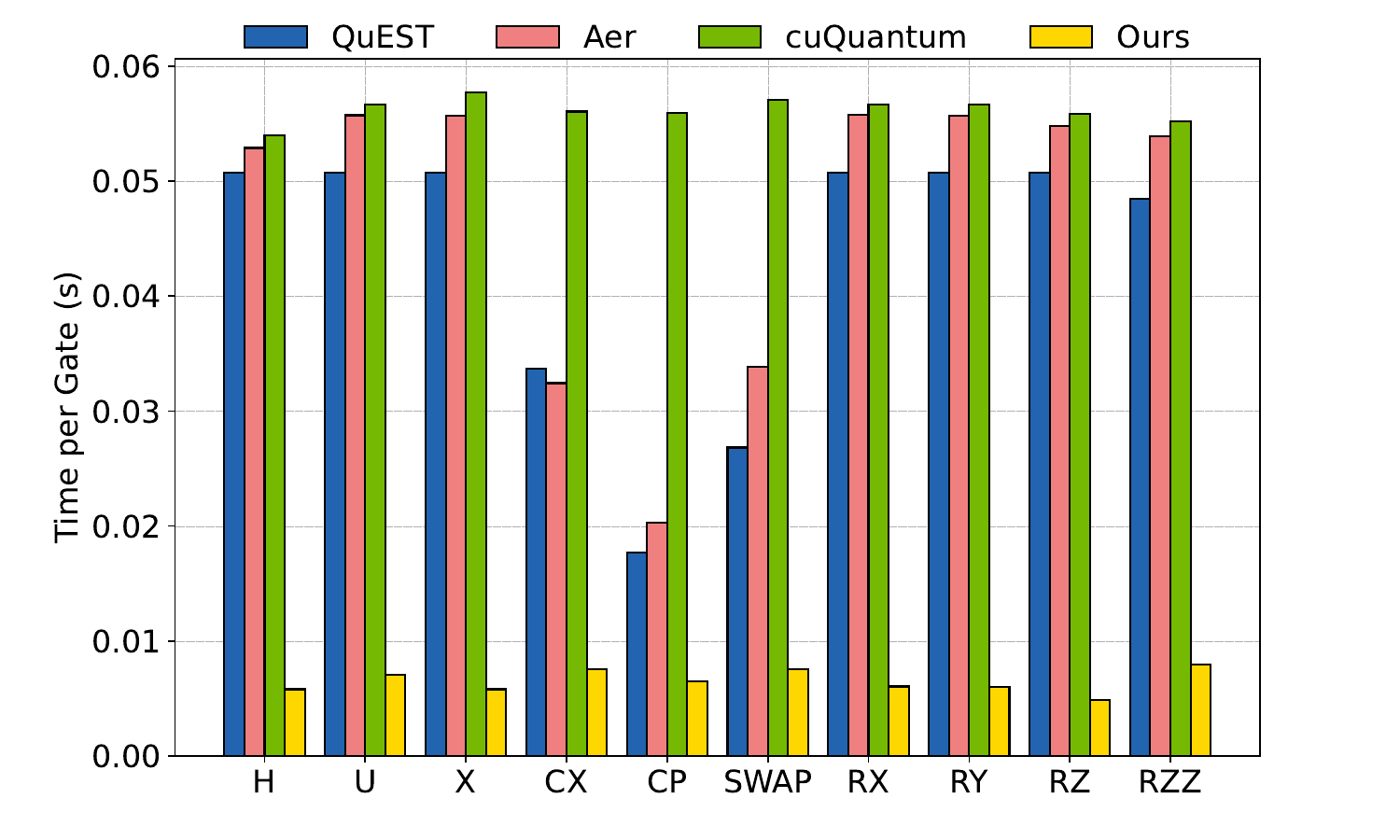}
\label{fig:gate_benchmark}
}\hfill
\subfloat[Circuit benchmark.]
{\includegraphics[width=0.43\textwidth]{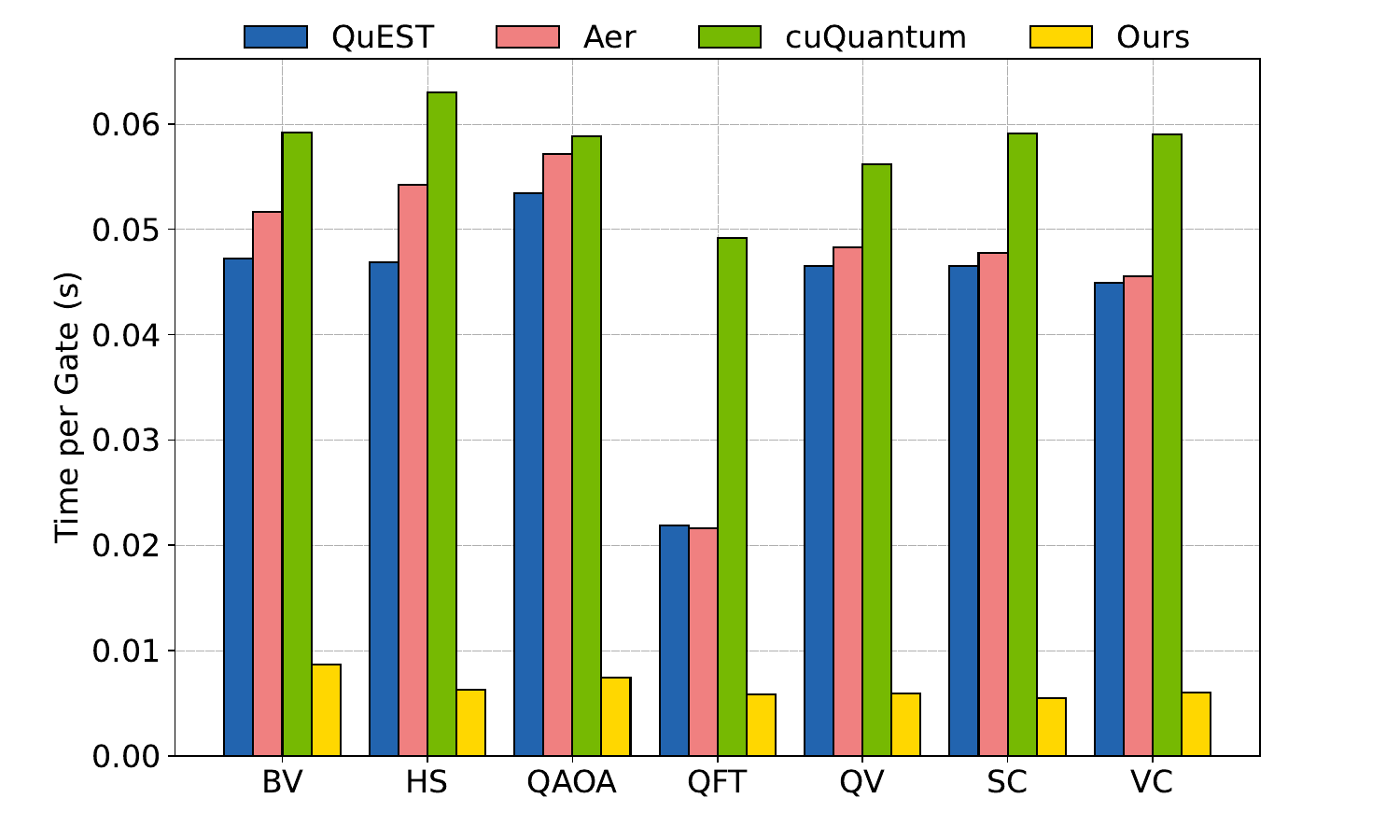}
\label{fig:circuit_benchmark}
}\\[-1ex]
\caption{Time per gate for 31-qubit gate and circuit benchmarks.}
\label{fig:all_benchmark}
\end{figure*}

\begin{figure*}[htb]
\captionsetup[subfloat]{farskip=2pt,captionskip=1pt}
\centering
\subfloat[Utilizing NVLink for XRS.]
{\includegraphics[width=0.45\textwidth]{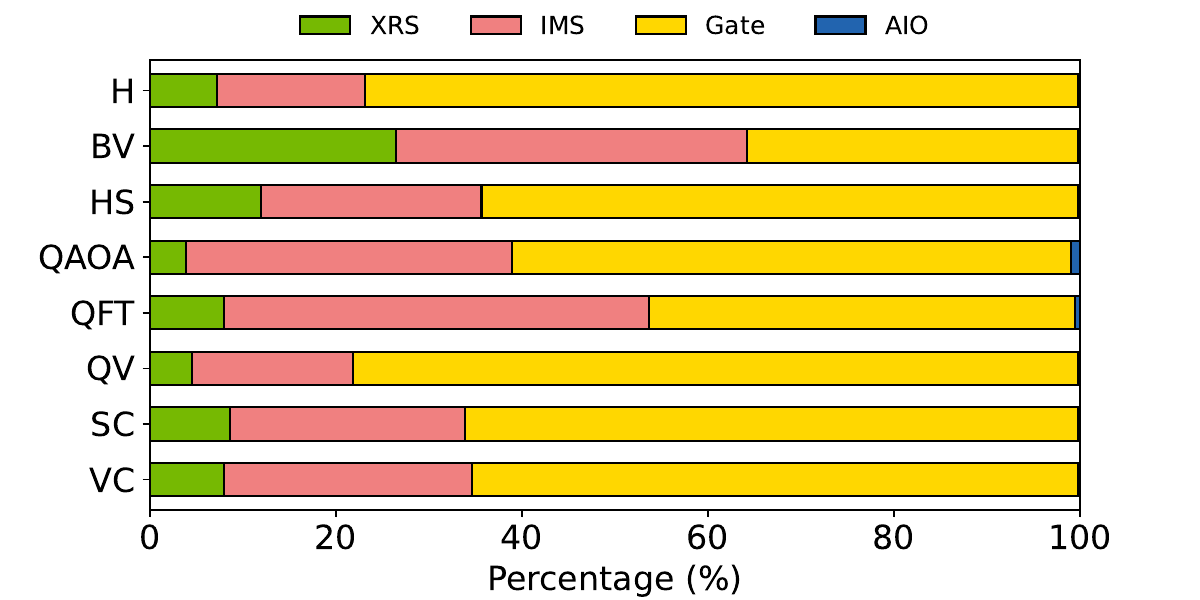}
\label{fig:ana}
}
\hfill
\subfloat[Utilizing sockets for XRS.]
{\includegraphics[width=0.45\textwidth]{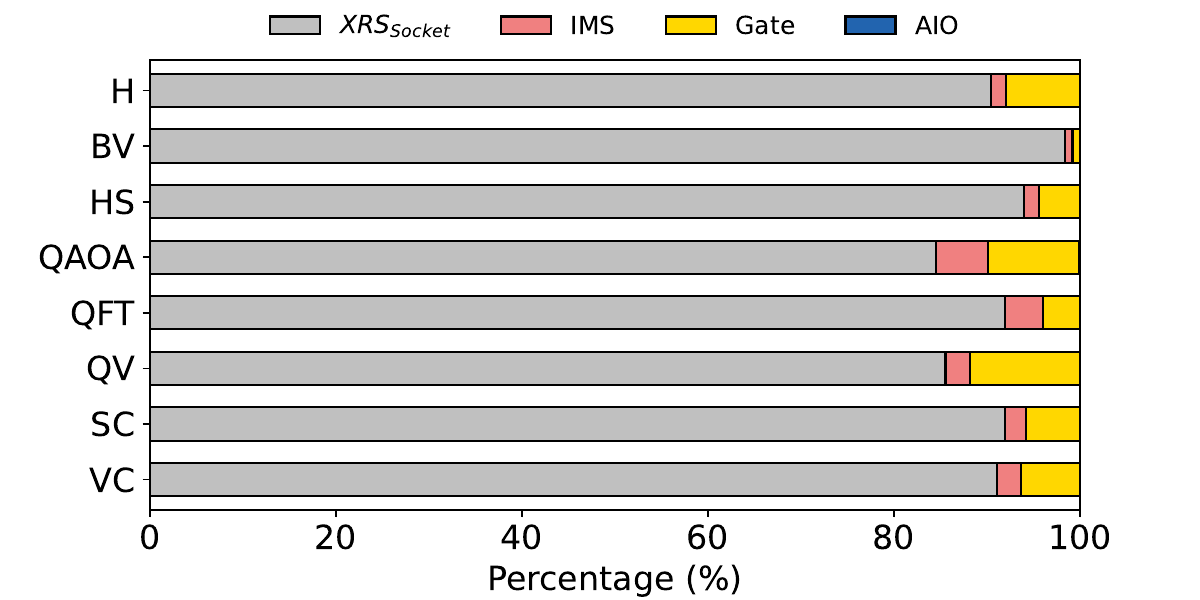}
\label{fig:ana_socket}
}
\\[-1ex]
\caption{Breakdown of 34-qubit simulation time for NVLink and socket as the backend of cross-rank swapping.}
\label{fig:ana_all}
\end{figure*}


\subsection{Qubit Benchmark and Strong Scaling}
\label{sec:h_benchmark}

\figurename{~\ref{fig:h_benchmark}}(a) demonstrates the simulation time of the H gate across a range from 24 to 31 qubits.
Most simulators exhibit consistent performance trends, and our simulator still emerges as the fastest in various qubit scenarios.
By Leveraging optimized gate block techniques and cache-friendly implementations, which facilitate data access in the GPU shared memory, the performance speedup can achieve a superior factor of 9.4 compared to cuQuantum in 31-qubit simulation.






To illustrate the superior speedup across multiple ranks in this most straightforward scenario, we also employ strong scaling metrics from 1 to 8 ranks exponentially \footnote{QuEST was omitted from the strong scaling evaluation due to its lack of support for simulating on multiple GPUs}.
Our benchmark results, as shown in~\figurename{~\ref{fig:h_benchmark}} (b), reveal an average speedup of approximately 8.7 times compared to other simulators.
The results confirm that the acceleration trend can also be proportional to the number of GPUs deployed for a simple gate benchmark, thereby affirming the robust scaling efficiency.





\subsection{Benchmark of Quantum Gates and Quantum Circuits}
\label{sec:gate_benchmark}
This section thoroughly explores the simulation efficiency for a variety of quantum gates and circuits on a single GPU, providing valuable insights for subsequent performance analysis and optimization.
As depicted in~\figurename{~\ref{fig:all_benchmark}} (a), our work outperforms all other simulators in all simulated gates, achieving an average speedup of 8.7 times over cuQuantum.

In detailed discussions, the Pauli-X gate can obtain a remarkable speedup of 9.9 times, benefiting from our series of performance tuning efforts, such as the removal of unnecessary floating-point computations.
Gate operations involving control qubits, such as CX and CPhase gates, benefit from matrix simplifications with halving computations, enabling effective optimizations by QuEST and Aer simulator. 
Implementing these performance tricks allows these types of gates to lag behind our simulator by only 4.3 and 2.7 times, respectively.

For the circuit-level benchmark, the specific simulation efficiency of up to seven circuits is demonstrated in \figurename{~\ref{fig:all_benchmark}} (b). 
Unsurprisingly, our performance results for these circuits once again comprehensively outperform other simulators, with a speedup of 9 times on average.
This achievement can be attributed to our optimal efficiency at the individual gate-level benchmark.
Regrettably, cuQuantum does not prioritize efficiency in these circuits with standard gate operations, resulting in comparatively lower performance.

\subsection{Performance Analysis}
\label{sec:performance_analysis}
To thoroughly pinpoint potential efficiency constraints, we dissect the performance attributes of our quantum simulator.
\figurename{~\ref{fig:ana_all}}(a) delineates the distribution of the entire simulation time across various processes within our simulator. The symbol~\emph{AIO} denotes the time spent executing the All-in-one optimization module during the preprocessing stage.
The~\emph{Gate} accounts for the duration allocated to the execution of quantum gate operations.
The~\emph{XRS} and~\emph{IMS} stand for time spans for the qubit swapping across rank and in-memory, respectively.

The most significant portion of the processing time is attributed to gate operations in~\figurename{~\ref{fig:ana_all}}(a) since they are the primary component for computation in our simulator.
Following gate operations, IMS emerges as the second most time-intensive component, representing a substantial investment in terms of execution time.
Incorporating this critical component enables optimized circuits to benefit from cache access properties, resulting in a remarkable speedup of up to 10 times.
This optimization still relies on the specific circuit.
In circuits where qubits have higher interdependence, such as in BV and QFT, more IMS consumption occurs, resulting in a performance enhancement of approximately 7-8 times.
The remaining processing time is allocated to two components, XRS and AIO.
Thanks to its polynomial-time algorithm, AIO remains time-efficient, even when executed exclusively through the CPU.
Leveraging the complete bandwidth of NVLink, XRS achieves minimal latency for data transfer.


When transmitting data without utilizing P2P (peer-to-peer) via NVLink and shared memory transports, the performance bottleneck shifts to XRS, resulting in a hundred-fold degradation compared to the NVLink approach, as illustrated in~\figurename{\ref{fig:ana_all}}(b).
In such a scenario, NCCL will utilize the network to communicate between the IP sockets, even when the GPUs are located on the same host. 
Moreover, due to IP sockets depending on the CPU for data copying and involving multiple data copies between user and kernel space, simulation efficiency is heavily limited.
This limitation becomes more significant when MPI-based simulators with IP sockets are used~\cite{mpiQulacs, QuEST, 45qubit}.

To solve this type of inter-communication issue, it is important to use the remote direct memory access (RDMA) technique for QCS~\cite{rdma_sim}.
Our simulator has also implemented the GPU-RDMA approach by partitioning the device into independently separate containers and the results precisely align with the following performance analysis.
In general, the bidirectional bandwidth of NVLink between any two GPUs in the DGX-A100 is 600GB/s. Suppose we connect multiple machines with InfiniBand fabric, and each machine has 8x 200Gbps InfiniBand NICs; then the inter-GPU communication bandwidth across machines would be 200Gbps on average, which is (200*2)/(600*8)=1/12 of NVlink.
Thus, the total execution time would be increased to 10\%*12+90\%=2.1 times, which means our MPI version would run at 47.6\% speed over InfiniBand, compared to NVLink.

In supercomputing environments, it is advisable to utilize the fastest communication techniques, despite the mitigating effects of GPU-RDMA on these issues.
With this in mind, NVIDIA GB200 NVL72~\cite{gh200_nvl72} featuring 5th generation NVLink technology (with bandwidth capabilities of 130 TB/s) emerges as a promising solution for QCS in a supercomputing environment.

\subsection{Roofline Model}
\label{sec:roofline_model}
To clarify the compute and memory impacts, we establish the roofline model for 3 NVIDIA GPUs with varying specifications, ranging from low to high, including RTX 3090, RTX 4090, and A100.
%
We utilized a standard of 30-qubit H-gate operations as the benchmark to conduct experiments under varying memory constraints.
Due to the identical type of gate operations, we employed the Nsight Compute profiler~\cite{nsight_doc} to sample a kernel function of core computations for a fair and concise presentation.
QuEST is selected as the representative simulation backend to simplify the experiments, and the term~\emph{GBG} denotes these simulators with gate-by-gate simulation schemes, such as Aer, cuQuantum, and Cirq qsim.

As depicted in \figurename{~\ref{fig:roofline_model}}, a clear trend from the lower left to the upper right is observed on all three GPUs, indicating that our simulations are universally effective across different GPU ranks.
For A100, the improvement in FLOPS can reach up to a factor of 8x due to its abundant floating-point computational units, while the overall arithmetic intensity experiences a dramatic increase of 96x as a result of our cache-aware implementations.
Surprisingly, our simulation performance on the RTX 3090 exceeds that of the GBG-based simulator on A100 by more than 1.13x.

It is worth noting that our simulator can be further accelerated with increased computing power. For instance, utilizing a next-generation GPU such as the GB200, which offers 90 TFLOPS compared to 9.7 TFLOPS for the A100, we expect a 9x speedup of the simulation. This substantial improvement would not be achievable with conventional memory-bound simulators~\cite{QuEST, Aer_sim, cuQuantum, Cirq} in the high-performance computing domain.

\begin{figure}[tb!]
\centerline{\includegraphics[width=.93\columnwidth]{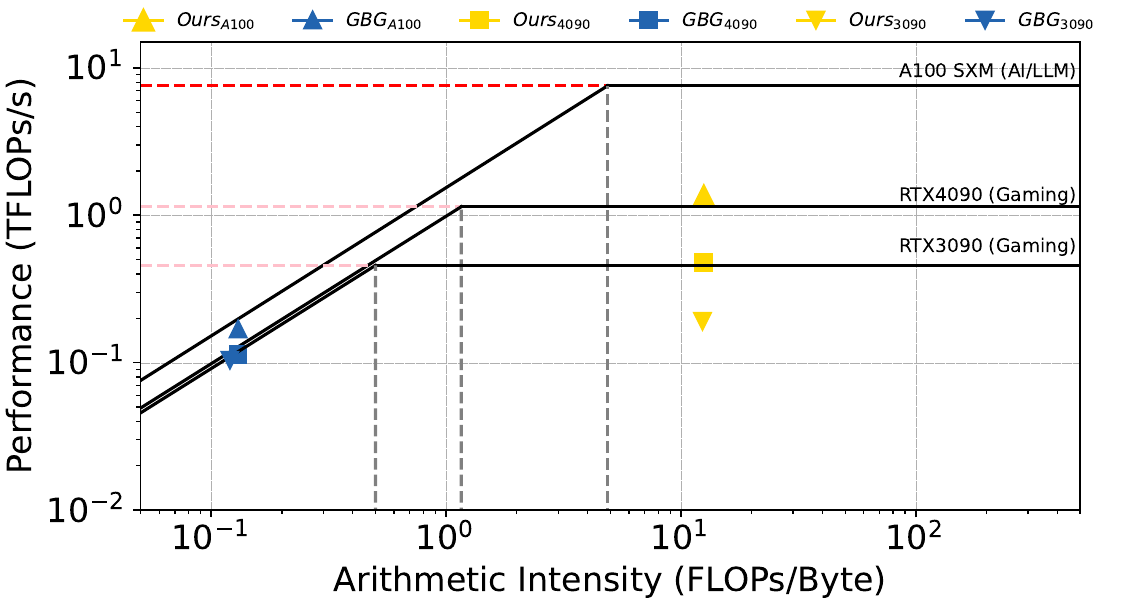}}
\caption{Roofline model of 30-qubit benchmark for 3 GPUs.}
\label{fig:roofline_model}
\end{figure}


\section{Conclusion}
\label{sec:conclusion}


In this paper, we present a versatile quantum circuit simulation toolkit to streamline the development and deployment of emerging quantum applications.
Designed primarily for supercomputing ranks, our simulation workflow boasts extensive parallelism and scalability.
Through the integration of the all-in-one optimization and all-in-cache simulation module, we effectively solve most performance bottlenecks encountered by existing high-performance state vector simulators. 
Compared to the state-of-the-art simulators, our toolkit achieves significant performance improvements averaging 9-fold times across various quantum gate and circuit benchmarks.
These encouraging results can be further validated through the performance profilers.
The breakdown of the simulation time demonstrates the superior transfer efficiency of NVLink, while the roofline model underscores our computational superiority.

While our simulator effectively addresses most critical performance issues in QCS, this area of research remains extremely challenging, especially in top-tier supercomputing conferences.
The lack of foundational knowledge in quantum circuit simulation and high-performance computing has led most literature to discuss only basic scalability on only CPU architectures without practical performance profiling and applying the RDMA techniques.
Most concerningly, many simulators have failed to achieve the performance reported in their papers or documents or even be irreproducible.
These phenomena have caused performance research in QCS to stagnate for over a decade.
Fortunately, your \textbf{`Queen'} has come.
We firmly believe that our contributions with strong reproducibility will unlock the complete potential of QCS.

In future work, we plan to extend this research to encompass the upcoming NVIDIA GB200 NVL72 for large-scale quantum circuit simulation while incorporating supplementary performance profiling techniques to pinpoint potential performance bottlenecks in increasingly intricate computation environments.

\bibliographystyle{ACM-Reference-Format}
\bibliography{ref}

\twocolumn[%
{\begin{center}
\Huge
Appendix: Artifact Description/Artifact Evaluation        
\end{center}}
]

\apppart{Reproducibility}
Unlike other claimed state-of-the-art simulators, this work can be reliably reproduced using the artifact description and artifact evaluation provided.
You are welcome to challenge our simulator performance or play with us.

\appendixAD

\section{Overview of Contributions and Artifacts}

\subsection{Paper's Main Contributions}



\begin{description}
\item [$C_1$] This paper introduces an all-in-one optimization module designed to convert original quantum circuits into optimized versions that are more tailored for execution on classical computers.
\item [$C_2$] This paper presents the implementation and performance analysis of the all-in-cache simulation module with a high-speed simulation workflow.
\item[$C_3$] This paper conducts a comparative analysis of simulation performance utilizing qubit, gate, and circuit benchmarks on multi-GPUs against state-of-the-art simulators.
\end{description}

\subsection{Computational Artifacts}



\textbf{For security purposes, to prevent unauthorized use, we kindly request that you contact the authors via email.}



\begin{center}
\begin{tabular}{rll}
\toprule
Artifact ID  &  Contributions &  Related \\
             &  Supported     &  Paper Elements \\
\midrule
$A_1$ & $C_1$ & Figure 5 \\
\midrule
$A_2$ & $C_2$, $C_3$ & Figure 8 \\
& & Figure 9 \\
& & Figure 10 \\
& & Figure 11 \\
& & Figure 12 \\
\bottomrule
\end{tabular}
\end{center}

\section{Artifact Identification}


\newartifact

\artrel


This artifact $A_1$ is an implementation of $C_1$, which is an optimization tool for doing the qubit reordering and gate fusion within the quantum circuits.

\artexp

This artifact aims to transform any circuits (e.g., QFT, QAOA, or BV) into the optimized circuit file, which includes gate blocks, in-memory swapping (SQS), and cross-rank swapping (CSQS).
Specifically, the input format for the circuit is structured as follows.
\begin{center}
\begin{tcolorbox}[breakable]
\textbf{[Gate type] \textcolor{red}{[Target qubits]} [ID] [Remain info]}
\end{tcolorbox}
\end{center}

Given an example circuit represented in Figure 5 (a), along with our specified input format, the optimization module is capable of producing the optimized circuit as output.
\begin{center}
\begin{tcolorbox}[breakable]
\textbf{H \textcolor{red}{0} 0}
\\\textbf{ H \textcolor{red}{1} 1}
\\\textbf{ RZZ \textcolor{red}{2 4} 2}
\\\textbf{ RZZ \textcolor{red}{5 7} 3}
\\\textbf{ H \textcolor{red}{8} 4}
\\\textbf{ H \textcolor{red}{9} 5}
\\\textbf{ H \textcolor{red}{3} 6}
\\\textbf{ H \textcolor{red}{6} 7}
\\\textbf{ RZZ \textcolor{red}{0 2} 8}
\\\textbf{ RZZ \textcolor{red}{4 7} 9}
\\\textbf{ H \textcolor{red}{9} 10}
\\\textbf{ RZZ \textcolor{red}{1 8} 11}
\\\textbf{ RZZ \textcolor{red}{3 6} 12}
\\\textbf{ H \textcolor{red}{5} 13}
\end{tcolorbox}
\end{center}

The output resulting from the qubit reordering optimization should match Figure 5(b) as follows.
The target qubit for each original gate operation should not exceed the chunk size of 4, which is also marked in red.
\begin{center}
\begin{tcolorbox}[breakable]
\textbf{3} \textcolor{ForestGreen}{\# Gate Block Size}
\\\textbf{H \textcolor{red}{0} 0}
\\\textbf{H \textcolor{red}{1} 1}
\\\textbf{H \textcolor{red}{3} 6}
\\\textbf{1}
\\\textbf{SQS 3 0 1 3 4 5 7} \textcolor{ForestGreen}{\# In-memory swapping}
\\\textbf{4}
\\\textbf{RZZ \textcolor{red}{0 2} 2}
\\\textbf{RZZ \textcolor{red}{1 3} 3}
\\\textbf{RZZ \textcolor{red}{0 3} 9}
\\\textbf{H \textcolor{red}{1} 13}
\\\textbf{1}
\\\textbf{SQS 3 0 1 3 4 6 7}
\\\textbf{3}
\\\textbf{H \textcolor{red}{1} 7}
\\\textbf{RZZ \textcolor{red}{1 3} 12}
\\\textbf{RZZ \textcolor{red}{0 2} 8}
\\\textbf{1}
\\\textbf{SQS 2 0 2 6 7}
\\\textbf{1}
\\\textbf{CSQS 2 6 7 8 9} \textcolor{ForestGreen}{\# Cross-rank swapping}
\\\textbf{1}
\\\textbf{SQS 2 0 2 6 7}
\\\textbf{1}
\\\textbf{SQS 1 3 5}
\\\textbf{4}
\\\textbf{H \textcolor{red}{2} 5}
\\\textbf{H \textcolor{red}{2} 10}
\\\textbf{H \textcolor{red}{0} 4}
\\\textbf{RZZ \textcolor{red}{0 3} 11}
\end{tcolorbox}
\end{center}

After applying all features in this module, the result should match Figure 5(c). The diagonal gates, such as RZZ, within each gate block, will be merged without altering correctness.
To avoid excessive elaboration, detailed results are not provided.
Please verify this matter effortlessly through the scripts we provided.

To verify the optimized circuit executes in the correct sequence and with proper dependencies
on target qubits, one can utilize our validation script as follows.
\begin{center}
\begin{tcolorbox}[breakable,colback=blue!5!white,colframe=blue!75!black]
\textbf{\$ }\textcolor{ForestGreen}{\# switch workspace to the validatio directory}
\\
\textbf{\$ cd ./tests}
\\
\textbf{\$ }\textcolor{ForestGreen}{\# run the validation script}
\\
\textbf{\$ ./validate\_circuit\_order.sh}
\end{tcolorbox}
\end{center}

If all optimized circuits have the correct sequence, the message "Passed all circuit order validations" will be output to the terminal. Otherwise, a mismatch error will be invoked.

\arttime


The expected results of this module are listed as follows.
\begin{easylist}
& The required time for Artifact Setup is negligible if our provided raw circuit file is adopted. Otherwise, installing the Python package of the open-source project may take 5-10 minutes.
& For Artifact Execution, each optimized circuit of our benchmark is consistently generated within 0.5 seconds.
& For Artifact Analysis, the optimized circuit does not represent the actual simulation efficiency, so we refrain from discussing it for now.
\end{easylist}

\artin

\artinpart{Hardware}

No specific Hardware is required to execute the experiments.

\artinpart{Operating System}
We employ Docker with Ubuntu 22.04 and kernel version 5.15.0-1044-nvidia for our experiments. 
However, no specific operating system is required to execute the experiments.

\artinpart{Software}

No specific software is required to execute the experiments.
However, for the convenience of downloading the open-source circuit generator and facilitating subsequent simulation stages, we still recommend installing all necessary environments at this stage.

\begin{easylist}
& In order to generate highly reproducible circuits in an open-source manner and facilitate direct comparisons with various works in one single execution environment, Python 3.10.12 should be used along with the following package dependencies. (i.e., using different configurations does not ensure running the experiments for all related works in one single environment):

    && attrs==20.3.0
    && cachetools==5.3.3
    && certifi==2021.5.30
    && charset-normalizer==3.3.2
    && cirq==0.13.1
    && cirq-aqt==0.13.1
    && cirq-core==0.13.1
    && cirq-google==0.13.1
    && cirq-ionq==0.13.1
    && cirq-pasqal==0.13.1
    && cirq-rigetti==0.13.1
    && cirq-web==0.13.1
    && contourpy==1.2.0
    && cuquantum-cu12==23.10.0
    && custatevec-cu12==1.5.0
    && cutensor-cu12==1.7.0
    && cutensornet-cu12==2.3.0
    && cycler==0.12.1
    && decorator==5.1.1
    && dill==0.3.8
    && duet==0.2.9
    && fonttools==4.49.0
    && google-api-core==1.34.1
    && google-auth==2.28.2
    && googleapis-common-protos==1.63.0
    && grpcio==1.62.1
    && grpcio-status==1.48.2
    && h11==0.9.0
    && httpcore==0.11.1
    && httpx==0.15.5
    && idna==2.10
    && iso8601==0.1.16
    && kiwisolver==1.4.5
    && lark==0.11.3
    && matplotlib==3.8.3
    && mpmath==1.3.0
    && msgpack==1.0.8
    && networkx==2.8.8
    && numpy==1.26.4
    && nvidia-cublas-cu12==12.3.4.1
    && nvidia-cuda-runtime-cu12==12.3.101
    && nvidia-cusolver-cu12==11.5.4.101
    && nvidia-cusparse-cu12==12.2.0.103
    && nvidia-nvjitlink-cu12==12.3.101
    && packaging==24.0
    && pandas==2.2.1
    && pbr==6.0.0
    && pillow==10.2.0
    && ply==3.11
    && protobuf==3.20.3
    && psutil==5.9.8
    && py==1.11.0
    && pyasn1==0.5.1
    && pyasn1-modules==0.3.0
    && pydantic==1.8.2
    && PyJWT==1.7.1
    && pyparsing==3.1.2
    && pyquil==3.0.1
    && python-dateutil==2.8.2
    && python-rapidjson==1.16
    && pytz==2024.1
    && pyzmq==25.1.2
    && qcs-api-client==0.8.0
    && qiskit==0.46.0
    && qiskit-aer-gpu==0.13.3
    && qiskit-terra==0.46.0
    && requests==2.31.0
    && retry==0.9.2
    && retrying==1.3.4
    && rfc3339==6.2
    && rfc3986==1.5.0
    && rpcq==3.11.0
    && rsa==4.9
    && ruamel.yaml==0.18.6
    && ruamel.yaml.clib==0.2.8
    && rustworkx==0.14.2
    && scipy==1.12.0
    && six==1.16.0
    && sniffio==1.2.0
    && sortedcontainers==2.4.0
    && stevedore==5.2.0
    && symengine==0.11.0
    && sympy==1.12
    && toml==0.10.2
    && tqdm==4.66.2
    && typing\_extensions==4.10.0
    && tzdata==2024.1
    && urllib3==2.2.1
\end{easylist}

\artinpart{Datasets / Inputs}


To lessen the load on users for recompilation, the experimental inputs consist of a designated configuration file and a total of 17 distinct gates and circuits, accessible via the ./circuit directory.
Alternatively, one can also generate all the raw circuits with the following command.

\begin{center}
\begin{tcolorbox}[breakable,colback=blue!5!white,colframe=blue!100!black]
\textbf{\$ }\textcolor{ForestGreen}{\# install required packages listed in the \emph{Software} section}
\\
\textbf{\$ pip3 install -r requirements.txt}
\\
\textbf{\$ }\textcolor{ForestGreen}{\# switch workspace to the following directory}
\\
\textbf{\$ cd ./optimizer}
\\
\textbf{\$ }\textcolor{ForestGreen}{\# generate raw circuits (e.g., BV, QAOA, ..., etc) }
\\
\textbf{\$ ./gen\_raw\_circuits.sh}
\\
\end{tcolorbox}
\end{center}


\artinpart{Installation and Deployment}
The version of the GNU g++ compiler used is 11.04.



\artcomp




A workflow may consist of two tasks: $T_1$ and $T_2$.
Task $T_1$ compiles the C++ source code for the all-in-one module.
Task $T_2$ leverages the module to transform the raw circuits in all experiments into optimized circuits.
The dependency of tasks is $T_1 \rightarrow T_2$
Note that the raw circuits have already been obtained in the \emph{dataset/input} section.



\artout
In terms of efficiency, the elapsed time for transforming any raw circuit is less than 1 second on a DGX-A100, thanks to the polynomial-time complexity of the implementation.

Moreover, this artifact demonstrates promising results in reducing the number of gate blocks.
For example, the 31-qubit QFT circuit with 496 gates is grouped into only 18 gate blocks.
Theoretically, memory access time could be reduced by a factor of 27 (496/18). However, in practice, this reduction is usually unattainable due to considerations involving compiler optimizations, making it more of a reference point.

The reductions in gate blocks result in fewer in-memory swaps and cross-rank swaps compared to other works, thereby enhancing an 8-fold times speedup while running the circuit through our simulation module ($A_2$) on the DGX-A100, a significance that is elaborated upon in more depth in the AE.

\newartifact

\artrel

This artifact $A_2$ is a comprehensive tool for quantum circuit simulation that orchestrates the complete implementation of $C_2$.
The integration of various benchmarks and other simulators can showcase the efficiency of $C_3$ (i.e., demonstrating higher efficiency compared to the other works).
Alongside the generated circuit via artifact $A_1$, this stage enables the presentation of a comprehensive performance report.
\artexp

The expected results of each benchmark on a DGX-A100 workstation are listed as follows.
These benchmarking results can dominate state vector-based simulation in various cases, as mentioned in 

\begin{easylist}
& QAOA Benchmark: \\Our work with fusion should get a performance improvement of 4.x compared to cuQuantum and aer with gate fusion, and our work in a single GPU without fusion still outperforms Aer and cuQuantum with fusion in 2 GPUs.
& QFT Benchmark: \\Our work with fusion should get a performance improvement of 4.x compared to cuQuantum and aer with gate fusion, and our work in a single GPU without fusion still outperforms Aer and cuQuantum with fusion in 2 GPUs.
& Qubit Benchmark: \\Our work should get an average performance improvement of 9.x compared to cuQuantum and Aer in various qubit scenarios.
& Strong Scaling Benchmark: \\This experiment demonstrates that our work has strong scaling capability, achieving an average performance improvement of 8.7x compared to cuQuantum and Aer in various rank scenarios.
& Gate Benchmark: \\ To show the versatility of various gates, our work should get an average performance improvement of 8.7x compared to cuQuantum and Aer.
& Circuit Benchmark: \\To show the versatility of various circuits, our work should get an average performance improvement of 9.x compared to cuQuantum and Aer.
& Breakdown of 34-qubit Simulation Time: \\Gate operation should occupy the most time. This experiment can be used to analyze the whole simulation process for different circuits.
& Breakdown of 34-qubit Simulation Time w/o NVLink and shared memory: \\The cross-rank swapping should occupy the most time. This experiment serves as a benchmark for comparison against NVLink and socket-based methodologies.
& Roofline Model of The 30-qubit H-gate Benchmark: \\Our work should be computation-bound, while QuEST should be memory-bound, highlighting the decisive difference that demonstrates us as the next-generation simulator.
\end{easylist}

\arttime

\begin{easylist}
& For Artifact Setup, the required time is negligible if our Dockerfile is adopted. 
Otherwise, installing the NVIDIA Driver, CUDA toolkit, and NCCL library may take up to 30 minutes, depending on network speed.

& For Artifact Execution, we provide the expected computational time of this artifact on a DGX-A100 as follows.
Notably, the proficiency with which information is gathered using a profiler varies depending on the familiarity with the operations.
    && QAOA Benchmark: 16 minutes
    && QFT Benchmark: 5 minutes
    && Qubit Benchmark: 4 minutes
    && Strong Scaling Benchmark: 3 minutes
    && Gate Benchmark: 15 minutes
    && Circuit Benchmark: 15 minutes
    && Breakdown of 34-qubit Simulation Time: 15 minutes
    && Breakdown of 34-qubit Simulation Time without NVLink: 150 minutes
    && Roofline Model: 3 minutes
& For Artifact Analysis, the required time is negligible when utilizing our provided script, which summarizes the results into a CSV and PDF file for each experiment.
\end{easylist}

\artin

\artinpart{Hardware}

For the experiments, we use a DGX-A100 workstation comprising two AMD EPYC 7742 CPUs and eight NVIDIA A100 SXM GPUs interconnected via NVLink, boasting a bandwidth of 600GB/s.

\artinpart{Operating System}
We employ Docker with Ubuntu 22.04 and kernel version 5.15.0-1044-nvidia for our experiments. 
However, no specific operating system is required to execute the experiments.

\artinpart{Software}
We use NVCC ver. 12.3.107 and the NCCL library ver. 2.20.3.
To reproduce the related works, the Python environment matches that of artifact $A_1$. You also need cmake to build QuEST, the related work we compared our work against. The version of cmake we use is ver. 3.22.1.

\artinpart{Datasets / Inputs}
All circuit inputs can be obtained after completing the steps of artifact $A_1$.

\artinpart{Installation and Deployment}

For compilation, we use NVCC ver. 12.3.107 with the "-lnccl -arch=sm\_80 -O3 
-{}- jump-table-density=0 -maxrregcount=64" flags and the NCCL library ver. 2.20.3 for compiling the CUDA codes.

To compile the executable file, execute "make -j" in the terminal within the ./simulator directory of the workspace path, and no specific requirements for deploying our work.

To execute the experiments, ensure the configure and circuit files exist, then run it with the following suggested command.
\begin{center}
\begin{tcolorbox}[breakable,colback=blue!5!white,colframe=blue!100!black]
\textbf{\$ ./simulator/Quokka -i [configure file] -c [circuit file]}
\end{tcolorbox}
\end{center}

\artcomp

A workflow may consist of three tasks: $T_1$, and $T_2$.
The task $T_1$ generates specific configure files, quantum gates, and circuits as benchmarks with optimizations for the simulators\footnote{The task $T_1$ represents the entire process of artifact $A_1$. If already completed, please skip this step.
}.
These configurations and benchmarks are then used as input by computational task $T_2$.
The outputs can be directly displayed in the terminal.
The results can also be automatically summarized as a CSV and PDF file if one utilizes our provided scripts.
The dependencies for reproducing our work are as follows: $T_1 \rightarrow T_2$.
The detailed step-by-step reproduction will be detailed in the Artifact Evaluation section.

The details of the experimental parameters are listed as follows.
\begin{easylist}
& In our simulator, specifying the size of the simulated qubits is achieved by switching the configuration file. This can be done by modifying the value of N in the .ini file.
& In our simulator, simulating different circuits only requires switching the .txt file without the need for recompilation.
Rest assured, our scripts, codes with comments, and README.md files offer abundant examples to aid in completing all experiments.
& The default number of qubits in the A100 scenario is set to 31 (this is the memory upper limit), but please adjust it according to your GPUs.
& There are 7 practical quantum circuits and 10 standard quantum gates used as benchmarks. 
You can still generate speciåfic simulation circuit files according to your needs, but please refer to the advanced user manual to complete this task.
& Given that the circuits executed in benchmarks and simulated scenarios are both large-scale, it is recommended to run the simulations multiple times to remove noise, even though running once is usually sufficient to indicate the simulation time.
\end{easylist}

\artout
Overall, we utilize the better elapsed time to identify a more effective simulation result.
A brief analysis and interpretation of each benchmark are listed below.
\begin{easylist}
& QAOA Benchmark:
\\This benchmark evaluates the overall simulation with gate fusion on a specific 31-qubit QAOA circuit.
Our work with gate fusion gets the fastest results on a DGX-A100.
& QFT Benchmark: 
\\This benchmark evaluates the overall simulation with gate fusion on a specific 31-qubit QFT circuit.
Our work with gate fusion gets the fastest results on a DGX-A100.
& Qubit Benchmark:
\\This benchmark evaluates the time per gate for the standard Hadamard gate with qubits ranging from 24 to 31.
It demonstrates the performance of the simulator under varying memory consumption levels.
Our work gets the fastest results for all cases on an A100.
& Strong scaling Benchmark:
\\This benchmark evaluates the time per gate for the standard Hadamard gate with ranks ranging from 1 to 8 exponentially.
It demonstrates the robust scalability of the simulator under varying GPU ranks.
Our work gets the fastest results for all cases on a DGX-A100.
& Gate benchmark:
\\This benchmark evaluates the time per gate for different types of gates on a 31-qubit system.
It demonstrates the high performance of each implementation for all varying gates.
Our work gets the fastest results for all cases on an A100.
& Circuit benchmark:
\\This benchmark evaluates the time per gate for different types of practical circuits on a 31-qubit system.
It demonstrates the high performance of each implementation for varying circuits.
Our work gets the fastest results for all cases on an A100.
& Breakdown of 34-qubit simulation time:
\\This hotspot analysis demonstrates the elapsed times of different techniques, providing insight into the performance bottlenecks of different models, with NVLink serving as the transfer medium for cross-rank communication.
Observing the changes in elapsed time, it can be noted that introducing in-memory swapping and cross-rank swapping is highly beneficial.
& Breakdown of 34-qubit simulation time without NVLink and shared memory:
\\This hotspot analysis demonstrates the elapsed times of different techniques, providing insight into the performance bottlenecks of different models without relying on NVLink for cross-rank communication. Observing the changes in elapsed time, it can be noted that utilizing NVLink for cross-rank swapping is highly beneficial.
& Roofline model of the 30-qubit H-gate benchmark: \
\\This analysis is typically used to determine whether the application is computation-bound or memory-bound.
By utilizing the Nsight Compute profiler, it is evident that our simulator is computation-bound across different levels of GPU.
For other simulators (we take QuEST as an example), the performance is significantly limited by memory bandwidth.
\end{easylist}



\appendixAE

\arteval{1}
\artin

To lessen the load on users for recompilation, the experimental inputs consist of a designated configuration file and a total of 17 distinct gates and circuits accessible via the ./circuit directory. 
Alternatively, one can also generate all the raw circuits with the following command
\begin{center}
\begin{tcolorbox}[breakable,colback=blue!5!white,colframe=blue!100!black]
\textbf{\$ }\textcolor{ForestGreen}{\# install required packages listed in the \emph{Software} section}
\\
\textbf{\$ pip3 install -r requirements.txt}
\\
\textbf{\$ }\textcolor{ForestGreen}{\# switch workspace to the following directory}
\\
\textbf{\$ cd ./optimizer}
\\
\textbf{\$ }\textcolor{ForestGreen}{\# download qasm.py from another github project.}
\\
\textbf{\$ }\textcolor{ForestGreen}{\# generate raw circuits (e.g., BV, QAOA, ..., etc) }
\\
\textbf{\$ ./gen\_raw\_circuits.sh}
\\
\textbf{\$ }\textcolor{ForestGreen}{\# compile the optimization module }
\\
\textbf{\$ make -j}
\end{tcolorbox}
\end{center}

\artcomp
Please switch the workspace to the optimizer directory and call the script to generate the gate blocks for the raw gate and circuit files.
\begin{center}
\begin{tcolorbox}[breakable,colback=blue!5!white,colframe=blue!100!black]
\textbf{\$ }\textcolor{ForestGreen}{\# switch workspace to the following directory}
\\
\textbf{\$ cd ./optimizer}
\\
\textbf{\$ }\textcolor{ForestGreen}{\# execution the optimization module for all raw circuits}
\\
\textbf{\$ ./gen\_gateblocks.sh }
\end{tcolorbox}
\end{center}

The compilation of the optimization module only requires using g++ to compile the finder.cpp file. The `gen\_gateblocks.sh' bash script iteratively optimizes and formalizes the raw circuits into the optimized circuits for our simulator via the gate block finding algorithm. The gates and circuits benchmark includes: ['bv', 'hs', 'qaoa', 'qft', 'qv', 'sc', 'vc', 'h', 'u', 'x', 'cx', 'cp', 'swap', 'rx', 'ry', 'rz', 'rzz'].

\artout


This artifact generates the necessary configuration and optimized circuit for the $A_2$ simulation module.
Although a comprehensive understanding of the optimization requires quantum computing expertise, we provide a simplified explanation.
Given the raw circuit with a specific chunk size, the target qubit of the output should fit within this region by inserting in-memory swapping and cross-rank swapping operations. 
This ensures that subsequent simulations can fully leverage the cache properties for quantum gate simulation.

The format for the circuit file should be as follows.
For detailed information, please refer to the provided README.md file.
\begin{center}
\begin{tcolorbox}[breakable]
\textbf{[Gate type] \textcolor{red}{[Target qubits]} [ID] [Remain info]}
\end{tcolorbox}
\end{center}

Using Figure 5(a) as an example, the raw circuit file is as follows.
\begin{center}
\begin{tcolorbox}[breakable]
\textbf{H \textcolor{red}{0} 0}
\\\textbf{ H \textcolor{red}{1} 1}
\\\textbf{ RZZ \textcolor{red}{2 4} 2}
\\\textbf{ RZZ \textcolor{red}{5 7} 3}
\\\textbf{ H \textcolor{red}{8} 4}
\\\textbf{ H \textcolor{red}{9} 5}
\\\textbf{ H \textcolor{red}{3} 6}
\\\textbf{ H \textcolor{red}{6} 7}
\\\textbf{ RZZ \textcolor{red}{0 2} 8}
\\\textbf{ RZZ \textcolor{red}{4 7} 9}
\\\textbf{ H \textcolor{red}{9} 10}
\\\textbf{ RZZ \textcolor{red}{1 8} 11}
\\\textbf{ RZZ \textcolor{red}{3 6} 12}
\\\textbf{ H \textcolor{red}{5} 13}
\end{tcolorbox}
\end{center}

Next, we need to understand how to apply our tool to create the correct format for our simulation module.
\begin{center}
\begin{tcolorbox}[breakable,colback=blue!5!white,colframe=blue!100!black]
\textbf{\$ ./optimizer/finder [target file] [cache size] [rank size] [qubit size] [apply in-memory swapping] [apply cross-rank swapping] [fusion size] [apply fusion technique]}
\end{tcolorbox}
\end{center}

\begin{easylist}
& The function of each parameter is listed as follows.
&& [target file]: The raw circuit file. 
&& [cache size]: The $2^{cache \ size}$ can not exceed the shared memory size of the GPU rank.
&& [rank size]: The $2^{rank \ size}$ can not exceed the primary memory size of the GPU rank.
&& [apply in-memory swapping]: Enable the in-memory swapping.
&& [apply cross-rank swapping]: Enable cross-rank swapping for multiple GPU environments.
&& [fusion size]: The $2^{fusion \ size}$ represents the region for gate fusion; this size should always not be larger than the cache size.
&& [apply fusion technique]: Enable gate fusion.
\end{easylist}

Now, input the following command to generate the optimized circuit as Figure5 (b).
\begin{center}
\begin{tcolorbox}[breakable,colback=blue!5!white,colframe=blue!100!black]
\textbf{\$ ./optimizer/finder ./circuit/example/raw.txt 4 8 10 1 1 0 0}
\end{tcolorbox}
\end{center}

The output should be as follows.
\begin{center}
\begin{tcolorbox}[breakable]
\textbf{3} \textcolor{ForestGreen}{\# Gate Block Size}
\\\textbf{H \textcolor{red}{0} 0}
\\\textbf{H \textcolor{red}{1} 1}
\\\textbf{H \textcolor{red}{3} 6}
\\\textbf{1}
\\\textbf{SQS 3 0 1 3 4 5 7} \textcolor{ForestGreen}{\# In-memory swapping}
\\\textbf{4}
\\\textbf{RZZ \textcolor{red}{0 2} 2}
\\\textbf{RZZ \textcolor{red}{1 3} 3}
\\\textbf{RZZ \textcolor{red}{0 3} 9}
\\\textbf{H \textcolor{red}{1} 13}
\\\textbf{1}
\\\textbf{SQS 3 0 1 3 4 6 7}
\\\textbf{3}
\\\textbf{H \textcolor{red}{1} 7}
\\\textbf{RZZ \textcolor{red}{1 3} 12}
\\\textbf{RZZ \textcolor{red}{0 2} 8}
\\\textbf{1}
\\\textbf{SQS 2 0 2 6 7}
\\\textbf{1}
\\\textbf{CSQS 2 6 7 8 9} \textcolor{ForestGreen}{\# Cross-rank swapping}
\\\textbf{1}
\\\textbf{SQS 2 0 2 6 7}
\\\textbf{1}
\\\textbf{SQS 1 3 5}
\\\textbf{4}
\\\textbf{H \textcolor{red}{2} 5}
\\\textbf{H \textcolor{red}{2} 10}
\\\textbf{H \textcolor{red}{0} 4}
\\\textbf{RZZ \textcolor{red}{0 3} 11}
\end{tcolorbox}
\end{center}

For the output, observe that the target qubit for each original gate operation should not exceed the chunk size of 4, as marked in red. Now, we continue by enabling the diagonal fusion function.
Please input the following command to generate the circuit files with all functions in this module.
\begin{center}
\begin{tcolorbox}[breakable,colback=blue!5!white,colframe=blue!100!black]
\textbf{\$ ./optimizer/finder ./circuit/example/raw.txt 4 8 10 1 1 4 1}
\end{tcolorbox}
\end{center}

The output should be as follows.
\begin{center}
\begin{tcolorbox}[breakable]
\textbf{3}
\\\textbf{H \textcolor{red}{0} 0}
\\\textbf{H \textcolor{red}{1} 1}
\\\textbf{H \textcolor{red}{3} 6}
\\\textbf{1}
\\\textbf{SQS 3 0 1 3 4 5 7}
\\\textbf{2}
\\\textbf{D4 \textcolor{red}{0 1 2 3} 0.75390225434330471 \emph{...}}
\\\textbf{H \textcolor{red}{1} 13}
\\\textbf{1}
\\\textbf{SQS 3 0 1 3 4 6 7}
\\\textbf{2}
\\\textbf{H \textcolor{red}{1} 7}
\\\textbf{D4 \textcolor{red}{0 1 2 3} -0.83907152907645244 \emph{...}}
\\\textbf{1}
\\\textbf{SQS 2 0 2 6 7}
\\\textbf{1}
\\\textbf{CSQS 2 6 7 8 9}
\\\textbf{1}
\\\textbf{SQS 2 0 2 6 7}
\\\textbf{1}
\\\textbf{SQS 1 3 5}
\\\textbf{4}
\\\textbf{H \textcolor{red}{0} 4}
\\\textbf{RZZ \textcolor{red}{0 3} 11}
\\\textbf{H \textcolor{red}{2} 5}
\\\textbf{H \textcolor{red}{2} 10}
\end{tcolorbox}
\end{center}

This resulting output should match Figure 5(c).
As mentioned above, you can also observe that the target qubit for each original gate operation should not exceed the chunk size of 4, as marked in red.
Moreover, the RZZ gates with IDs 2, 3, and 9 are now merged into a single D4 gate, and RZZ gates with IDs 7, 8, and 12 are also merged into a single D4 gate.
The symbol "\emph{...}" represents the remaining information of the fused gates.

Congratulations! You now have enough knowledge to discern the correctness of more complex circuits in our benchmark.
Of course, we also invite you to verify via our scripts.

\textbf{[Auto-validation]}
To validate that the optimized circuit executes in the correct sequence and with proper dependencies on target qubits, one can utilize our validation script. The processes are outlined as follows:
\begin{center}
\begin{tcolorbox}[breakable,colback=blue!5!white,colframe=blue!100!black]
\textbf{\$ }\textcolor{ForestGreen}{\# switch workspace to the validation directory}
\\
\textbf{\$ cd ./tests}
\\
\textbf{\$ }\textcolor{ForestGreen}{\# run the validation script}
\\
\textbf{\$ ./validate\_circuit\_order.sh}
\end{tcolorbox}
\end{center}

This script will invoke `dump\_order.py' for both the original circuit and the optimized one to perform detailed checking.
In summary, `dump\_order.py' traverses the circuit's operations on qubits and restores the optimized circuit to its original positions according to the sequence of swaps performed during optimization.
After the `diff' command in the script is used to compare the differences between the two circuits, the validation results will be provided.

If all pairs of circuits are identical, it indicates that the transformed circuit is correct.
In this case, the message "Passed all circuit order validations" will be output to the terminal.
Otherwise, the mismatch error will be invoked.

\arteval{2}

\artin

If you are using the Docker environment provided by us, you can skip this step.
Otherwise, please install NVCC version 12.3.107, NCCL library version 2.20.3, cmake 3.22.1 and the Python packages for the related works on your machine.
Nsight System v2023.2.3 is required to analyze the execution time. Other versions contain unexpected bugs from NVIDIA and may not work successfully.

\begin{easylist}
& To install the NVCC toolkit, please follow the instructions on the official website: \\ https://docs.nvidia.com/cuda/cuda-installation-guide-linux/index.html

& To install the NCCL library, please follow the instructions on GitHub: https://github.com/NVIDIA/nccl

& To install Nsight System v2023.2.3, please download and follow the instructions on the official website: https://developer.nvidia.com/gameworksdownload

& To install the Python 3.10.12 package for the related works, please follow the instructions as follows.
\begin{center}
\begin{tcolorbox}[breakable,colback=blue!5!white,colframe=blue!100!black]
\textbf{\$ }\textcolor{ForestGreen}{\# install required packages listed in \emph{Software} section}
\\
\textbf{\$ pip3 install -r requirements.txt}
\\
\textbf{\$ }\textcolor{ForestGreen}{\# switch workspace to the following directory}
\\
\textbf{\$ cd ./simulatior}
\\
\textbf{\$ }\textcolor{ForestGreen}{\# compile the simulation module }
\\
\textbf{\$ make -j}
\end{tcolorbox}
\end{center}
\end{easylist}

\artcomp

To run each benchmark, we now need to understand how to apply our tool to create
the correct format for our simulation module.
\begin{center}
\begin{tcolorbox}[breakable,colback=blue!5!white,colframe=blue!100!black]
\textbf{\$ ./simulator/Quokka -i [configure file] -c [circuit file]}
\end{tcolorbox}
\end{center}

\begin{easylist}
& The function of each parameter is listed as follows.
&& [configure file]: The configure file for the simulations.
&& [circuit file]: The circuit file for the simulations.
\end{easylist}

The example of the circuit file is the output of the optimizer, as mentioned above, and the example of the configure file with comments to explain each parameter is provided below.
\begin{center}
\begin{tcolorbox}
\textbf{[system]}
\\
\textbf{total\_qbit=31} \textcolor{ForestGreen}{// Total qubit for simulation}
\\
\textbf{rank\_qbit=0} \textcolor{ForestGreen}{ // The total ranks for running the simulation.}
\\
\textbf{buffer\_qbit=28} \textcolor{ForestGreen}{// The size of buffer for data transfer.}
\end{tcolorbox}
\end{center}

The example for running the specific circuit and configuration file is as follows:
\begin{center}
\begin{tcolorbox}[breakable,colback=blue!5!white,colframe=blue!100!black]
\textbf{\$ ./simulator/Quokka -i ./circuit/sub0/30/gpu.ini -c ./circuit/sub0/30/h30.txt}
\end{tcolorbox}
\end{center}

For the initial step, please switch the workspace to the `benchmarks' directory, and then build QuEST, the related work that we compared our work against.

\begin{center}
  \begin{tcolorbox}[breakable,colback=blue!5!white,colframe=blue!100!black]
\textbf{\$ }\textcolor{ForestGreen}{\# switch workspace to the following directory}
\\
\textbf{\$ cd ./bechmarks}
\\
\textbf{\$ }\textcolor{ForestGreen}{\# build QuEST simulator for benchmarks}
\\
\textbf{\$ ./build\_QuEST.sh}
  \end{tcolorbox}
\end{center}

The executable of QuEST will be built and located in `./benchmarks/QuEST/build/demo' relative to the artifact's root directory.

Then, to run the experiments and generate the figures automatically, as illustrated in the paper, please keep your working directory in `./benchmarks' and run the commands listed below.

\begin{easylist}    
  & QAOA Benchmark:
  \begin{center}
  \begin{tcolorbox}[breakable,colback=blue!5!white,colframe=blue!100!black]
  \textbf{\$ ./1a\_strong\_scale\_qaoa.sh $>$ qaoa.csv}
  \\
  \textbf{\$ python3 1\_strong\_scale\_circuit.py qaoa.csv qaoa.pdf}
  \end{tcolorbox}
  \end{center}
  && The first command will execute the QAOA simulation in our simulator, Aer simulation, and cuQuantum, while redirecting the results to the CSV file. The second command can generate a PDF file, as shown in Figure 8(a) in our paper.
  & QFT Benchmark: 
  \begin{center}
  \begin{tcolorbox}[breakable,colback=blue!5!white,colframe=blue!100!black]
  \textbf{\$ ./1b\_strong\_scale\_qft.sh $>$ qft.csv}
  \\
  \textbf{\$ python3 1\_strong\_scale\_circuit.py qft.csv qft.pdf}
  \end{tcolorbox}
  \end{center}
  && Similarly to the QAOA benchmark, the first command will execute the QFT simulation in all simulators while redirecting the results to the CSV file. The second command can generate a PDF file, as shown in Figure 8(b) in our paper.
  & Qubit Benchmark:
  \begin{center}
  \begin{tcolorbox}[breakable,colback=blue!5!white,colframe=blue!100!black]
  \textbf{\$ ./2a\_qubit\_benchmark.sh $>$ qubit.csv}
  \\
  \textbf{\$ python3 2a\_qubit\_benchmark.py qubit.csv qubit.pdf}
  \end{tcolorbox}
  \end{center}
  && The first command will execute the 31-qubit h-gate simulation in all simulators ranging from 24 to 30 qubits while redirecting the results to the CSV file. The second command can generate a PDF file, as shown in Figure 9(a) in our paper.
  & Strong scaling Benchmark:
  \begin{center}
  \begin{tcolorbox}[breakable,colback=blue!5!white,colframe=blue!100!black]
  \textbf{\$ ./2b\_strong\_scale.sh $>$ scaling.csv}
  \\
  \textbf{\$ python3 2b\_strong\_scale.py scaling.csv scaling.pdf}
  \end{tcolorbox}
  \end{center}
  && The first command will execute the strong scaling test for 31-qubit h-gate in all simulators while redirecting the results to the CSV file. The second command can generate a PDF file, as shown in Figure 9(b) in our paper.
  & Gate benchmark:
  \begin{center}
  \begin{tcolorbox}[breakable,colback=blue!5!white,colframe=blue!100!black]
  \textbf{\$ ./3a\_gate\_benchmark.sh $>$ gates.csv}
  \\
  \textbf{\$ python3 3a\_gate\_benchmark.py gates.csv gates.pdf}
  \end{tcolorbox}
  \end{center}
  && The first command will execute the strong scaling test for 31-qubit simulation of various types of gates in all simulators while redirecting the results to the CSV file. The second command can generate a PDF file, as shown in Figure 10(a) in our paper.
  & Circuit benchmark:
  \begin{center}
  \begin{tcolorbox}[breakable,colback=blue!5!white,colframe=blue!100!black]
  \textbf{\$ ./3b\_circuit\_benchmark.sh $>$ circuits.csv}
  \\
  \textbf{\$ python3 3b\_circuit\_benchmark.py circuits.csv circuits.pdf}
  \end{tcolorbox}
  \end{center}
  && The first command will execute the strong scaling test for 31-qubit simulation of various types of circuits in all simulators while redirecting the results to the CSV file. The second command can generate a PDF file, as shown in Figure 10(b) in our paper.
  & Breakdown of 34-qubit simulation time:
  \begin{center}
  \begin{tcolorbox}[breakable,colback=blue!5!white,colframe=blue!100!black]
  \textbf{\$ ./4a\_breakdown.sh $>$ nvlink.csv}
  \\
  \textbf{\$ python3 4a\_breakdown.py nvlink.csv nvlink.pdf}
  \end{tcolorbox}
  \end{center}
  && The first command will execute nsys profiler for all 34-qubit simulations of various types of circuits in all simulators while redirecting the results to the CSV file. The second command can generate a PDF file, as shown in Figure 11(a) in our paper.
  & Breakdown of 34-qubit simulation time without NVLink and shared memory:
  \begin{center}
  \begin{tcolorbox}[breakable,colback=blue!5!white,colframe=blue!100!black]
  \textbf{\$ ./4b\_breakdown.sh $>$ socket.csv}
  \\
  \textbf{\$ python3 4b\_breakdown.py socket.csv socket.pdf}
  \end{tcolorbox}
  \end{center}
  && The first command will execute nsys profiler for all 34-qubit simulations of various types of circuits in all simulators without applying NVLink and shared memory while redirecting the results to the CSV file. The second command can generate a PDF file, as shown in Figure 11(b) in our paper.
  \\To disable NVLink and shared memory support, the script includes specific parameters for the nccl library, such as `NCCL\_P2P\_DISABLE=1 and NCCL\_SHM\_DISABLE=1'.
  & Roofline model:
  \begin{center}
  \begin{tcolorbox}[breakable,colback=blue!5!white,colframe=blue!100!black]
  \textbf{\$ ./5\_roofline.sh $>$ roofline.csv}
  \\
  \textbf{\$ python3 5\_roofline.py roofline.csv roofline.pdf}
  \end{tcolorbox}
  \end{center}
  && The first command will execute the NSight compute profiler for all 30-qubit h-gate simulations in all simulators while redirecting the results to the CSV file. The second command can generate a PDF file, as shown in Figure 12 in our paper.
  \\ In general, this script only generates result for a single rank. To include additional ranks, such as RTX 4090 and 3090, please adhere to our guidelines and manually input their specifications into the script.
\end{easylist}

\artout
The figures and tables in our experimental results can be automatically produced by our script in the previous subsection, and the analysis for each experiment is listed below.
\begin{easylist}
  & QAOA and QFT Benchmark:
  \\ The strong scaling test for these two typical benchmarks comprehensively demonstrates our achievements in surpassing other simulators across various metrics. From one to eight GPUs, with or without diagonal fusion enabled, all our experiments have consistently outperformed other simulators with an average speedup of 8x times.
  & Qubit Benchmark and Strong Scaling Benchmark:
  \\ This experiment serves to support a more detailed comparison. We utilize the standard 31-qubit H-gate at this stage and validate its effects across different qubits and under strong scaling tests. These results indicate that our simulator achieves the best 9x acceleration and demonstrates excellent scalability.
  & Gate benchmark and Circuit benchmark:
  This experiment aims to demonstrate that our implementation techniques can be applied across various scenarios, such as standard gates and circuits. Among them, cuQuantum exhibits the poorest simulation performance, while we boast the best execution efficiency. The fundamental reason lies in our specific optimizations for each gate rather than relying solely on transpilation techniques in IBM-qiskit preprocessing. On DGX-A100, our related improvement in efficiency (i.e., time per gate) ranges between 5-10x.
  & Breakdown of 34-qubit simulation time with and without NVLink:
  \\ We profile our simulation at this stage using the NVIDIA nsys profiler. The results from the profiler should be divided by the total number of machines, as the measured data represents the total time of all machines. The elapsed time for AIO optimization is measured separately, but since it typically consumes less than 0.1\% of the time, it is not merged into the profiler to reduce the programming overhead. Primarily, despite the introduction of numerous swap operations, the majority of the time is dedicated to gate operations. 
  Additionally, when NVLink and shared memory on the host are disabled, there is a notable increase in data transfer time.
  This observation indicates that socket-based approaches may not be entirely optimal for supercomputing environments. Therefore, the approaches that offer high bandwidth and low latency are necessary, exemplified by the latest NVLink technology.
  & Roofline model:
  \\ In order to fairly compare the optimized results, all reports of the roofline model are obtained by the Nsight Compute profiler without additional manual processing. The result clearly shows that our simulator is computation-bound, and the other is the memory-bound program.
  We have overwhelmingly superior performance in both TFLOPS (8x) and Arithmetic Intensity (96x) on DGX-A100.
   We also demonstrate this performance profiling on RTX-3090 and RTX-4090. 
   The results indicate that our performance in FLOPS on RTX-3090 surpasses that of QuEST on A100.
\end{easylist}

\subsection{Summary}
\textbf{Queen}: ``My little princesses and princes, the other self-proclaimed state-of-the-art works are either irreproducible or indeed quite slow. 
Until when should this truth be concealed?
I'm so bored. Would you play with me?"

\end{document}